\documentclass[groupedaddress,superscriptaddress,amsmath,amssymb,pra]{revtex4}
\usepackage{url}
\usepackage{bm}
\usepackage{graphicx,graphics}
\usepackage{soul}
\usepackage{cancel}

\usepackage{algcompatible}
\usepackage{newfloat}

\DeclareFloatingEnvironment[
    fileext=loa,
    listname=List of Algorithms,
    name=ALGORITHM,
    placement=tbhp,
]{algorithm}

\usepackage{braket}
\usepackage[normalem]{ulem} 
\usepackage{xcolor} 
\newtheorem{definition}{Definition}

\usepackage{amsmath}
\newcommand{\bS}{\mathbb{S}}
\newcommand{\bM}{\cal{M}}

\newcommand{\w}[1]{\widetilde{#1}}
\newcommand{\low}[1]{\lfloor #1 \rfloor}
\newcommand{\half}[1]{(#1)_\frac{1}{2}}
\newcommand{\diag}[1]{\mathbb D(#1)}

\DeclareMathOperator*{\Max}{maximize}
\DeclareMathOperator*{\Min}{minimize}

\begin{document}

\title{Riemannian geometry and automatic differentiation for optimization problems \\ of quantum physics and quantum technologies}

\author{Ilia A. Luchnikov}
\affiliation{Moscow Institute of Physics and Technology,
Institutskii Pereulok 9, Dolgoprudny, Moscow Region 141700,
Russia}
\affiliation{Skolkovo Institute of Science and Technology, Skolkovo, Moscow Region 121205, Russia}
\affiliation{Russian Quantum Center, Skolkovo, Moscow 143025, Russia}

\author{Mikhail E. Krechetov} \affiliation{Skolkovo Institute of Science and Technology, Skolkovo, Moscow Region 121205, Russia}

\author{Sergey N. Filippov}
\affiliation{Moscow Institute of Physics and Technology,
Institutskii Pereulok 9, Dolgoprudny, Moscow Region 141700,
Russia} \affiliation{Steklov Mathematical Institute of
Russian Academy of Sciences, Gubkina Street 8, Moscow 119991,
Russia} \affiliation{Valiev Institute of Physics and Technology of Russian Academy of Sciences, Nakhimovskii Prospect 34, Moscow 117218, Russia}

\begin{abstract}
Optimization with constraints is a typical problem in quantum physics and quantum information science that becomes especially challenging for high-dimensional systems and complex architectures like tensor networks. Here we use ideas of Riemannian geometry to perform optimization on the manifolds of unitary and isometric matrices as well as the cone of positive-definite matrices. Combining this approach with the up-to-date computational methods of automatic differentiation, we demonstrate the efficacy of the Riemannian optimization in the study of the low-energy spectrum and eigenstates of multipartite Hamiltonians, variational search of a tensor network in the form of the multiscale entanglement-renormalization ansatz, preparation of arbitrary states (including highly entangled ones) in the circuit implementation of quantum computation, decomposition of quantum gates, and tomography of quantum states. Universality of the developed approach together with the provided open source software enable one to apply the Riemannian optimization to complex quantum architectures well beyond the listed problems, for instance, to the optimal control of noisy quantum systems.
\end{abstract}

\maketitle

\textit{Note added}. After the paper was published, Alexander
Pechen brought relevant references to our attention in October
2021: in Ref.~\cite{pechen-2008} the authors proposed an idea to
use complex Stiefel manifolds for parameterizing quantum channels
and performing the gradient optimization for quantum control and
quantum technologies; in Ref.~\cite{oza-2009} the authors
developed an approach to optimization of quantum systems with an
arbitrary finite dimension for quantum control and quantum
technologies via gradient flows over complex Stiefel manifolds.

\section{Introduction}
Ideas of modern deep learning~\cite{krizhevsky2012imagenet, goodfellow2014generative, vaswani2017attention, goodfellow2016deep, lecun2015deep, kingma2013auto, kingma2014adam} have been spreading toward various sciences~\cite{carleo2019machine, webb2018deep, goh2017deep}. Deep learning techniques have been successfully applied to different branches of physics  from statistical~\cite{wu2019solving, li2018neural, torlai2016learning, efthymiou2019super} and quantum mechanics~\cite{carleo2017solving, sharir2020deep, choo2019two, carrasquilla2019reconstructing, carrasquilla2017machine, luchnikov2020machine, luchnikov2019variational, torlai2019integrating, torlai2018neural} to nonlinear dynamical systems~\cite{raissi2019physics, lusch2018deep, li2019learning} and fluid dynamics~\cite{king2018deep, kutz2017deep, portwood2019turbulence}. Essentially, deep learning is a gradient-based optimization of complex computational models such as neural networks, tensor networks \cite{liao2019differentiable, pan2019contracting, hasik2019towards, torlai2019wavefunction}, solvers for ordinary and partial differential equations~\cite{chen2018neural, wang2020differentiable, toth2019hamiltonian, holl2020learning}, and Monte Carlo simulation schemes \cite{zhang2019automatic, schulman2015gradient, jang2016categorical}. The gradient-based optimization for the above models often implies the use of an \emph{automatic differentiation} algorithm, e.g., one of those reviewed in Ref.~\cite{baydin2017automatic}. Automatic differentiation makes the computation process differentiable with respect to its constituent elements (e.g., hidden layers in deep learning). Therefore, computation processes with automatic differentiation are readily tunable via gradient-based methods, which enables one to find a response of the solution to an external perturbation of the problem parameters (e.g., initial state in dynamical problems) and facilitates optimization of the solution properties (e.g., state energy in static problems).

In quantum physics problems, however, the elements to be optimized are usually restricted to some class. For instance, in circuit implementation of quantum computation, the elementary operations are to be unitary. Similarly, tensor network representations of multipartite quantum states impose restrictions on constituent tensors~\cite{orus2019tensor}. Finally, a general quantum state is given by a density matrix, i.e., a positive-semidefinite matrix with unit trace. Any valid optimization over such computation elements must respect the corresponding restrictions, which poses an important problem of how to perform optimization on specific manifolds (e.g., a manifold of unitary or isometric matrices). This is the Riemannian optimization~\cite{absil2009optimization, becigneul2018riemannian, li2020efficient, edelman1998geometry, tagare2011notes} that adapts the gradient-based methods in such a way that the incremented element still satisfies the desired property (unitarity, isometry, positive semidefiniteness, etc.). For instance, the Riemannian optimization with the orthogonality constraint $K^T K = KK^T = I$ on the real kernel $K$ of a recurrent neural network has been used to avoid the gradient explosion or vanishing in Refs.~\cite{lezcano2019cheap, vorontsov2017orthogonality}.

In the present work, we develop the first-order Riemannian
optimization methods over corresponding manifolds and combine them
with automatic differentiation to solve various problems of
quantum physics. Working with the manifold of unitary matrices, we
demonstrate how to prepare an arbitrary multiqubit quantum state
(including a highly entangled one) or a gate via circuit quantum
computation. Working with the manifold of isometric matrices
(complex Stiefel manifold), we compare the first-order Riemannian
optimization methods with an algorithm that is standard for
optimization of isometric tensor networks such as multiscale
entanglement-renormalization ansatz (MERA)~\cite{vidal2008class}
and unitary hierarchical tensor networks~\cite{liu2019machine}. In
the case of a local Hamiltonian for a multiqubit system, we also
perform optimization over a tensor network in the form of MERA and
find the low-lying spectrum of a Hamiltonian with no need to
explicitly calculate the reduced states and environments thanks to
the automatic differentiation. Working with the cone of positive
definite matrices, we derive the Riemannian gradient for two types
of metric. We compare the different first-order Riemannian
optimization methods by solving the illustrative problem of the
ground state search and apply these methods to tomography of
quantum states via the maximum likelihood estimation. Our findings
demonstrate that the Riemannian optimization and the automatic
differentiation form a new powerful numerical tool for solving
timely problems of quantum technologies. We also provide an
open-source library~\cite{QGOpt_repo} that allows performing
optimization with many natural ``quantum" constraints including
the constraints of positive-definiteness and isometry considered
in this paper.

The paper is organized as follows. In Section~\ref{section-RG}, we overview some basic machinery of Riemannian geometry needed in solving optimization problems. In Section~\ref{section-RG-in-QM}, we discuss some examples of the Riemannian manifolds emerging in the fields of quantum physics and quantum information science. In Section~\ref{section-RO}, we highlight the concept of the Riemannian gradient and describe how to restrict first-order optimization methods to a Riemannian manifold. In Section~\ref{section-AD-and-RO}, we overview the automatic differentiation algorithm. In Section~\ref{section-Stiefel}, we overview known methods for the Riemannian optimization on the complex Stiefel manifolds of unitary and isometric matrices. In Section~\ref{section-Hamiltonian-renorm}, we compare those methods with the conventional algorithm for optimization of isometric tensor networks. In Sections~\ref{section-entangling}, \ref{section-state-preparation}, and \ref{section-design}, we combine those methods with automatic differentiation to solve some problems of quantum control (aimed at creating a desired multiqubit entangled state or decomposing a quantum gate). In Section~\ref{section-ER}, we combine the Riemannian optimization methods on the complex Stiefel manifolds with the automatic differentiation to search for a ground state of local Hamiltonians via MERA tensor network. In Section~\ref{section-cone-Spp}, we consider the Riemannian geometry for a cone of positive definite matrices and derive the Riemannian gradients for two types of metric, which enables us to perform the Riemannian optimization over general quantum states and solve different optimization problems with positivity constraints, e.g., the tomography of quantum states. Finally, we make conclusions in Section~\ref{section-concl}.

\section{Riemannian geometry} \label{section-RG}
\begin{figure}
    \centering
    \includegraphics[width=15cm]{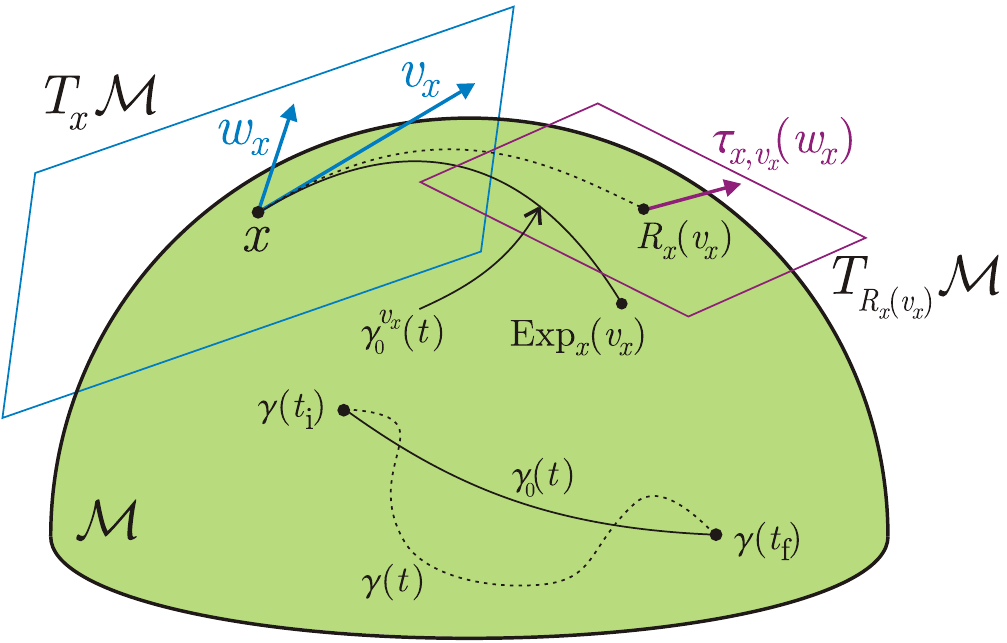}
\caption{Basic notions of Riemannian geometry. ${\cal M}$ is a
Riemannian manifold. A smooth mapping $\gamma$ from an interval
$(t_{\rm i},t_{\rm f})$ in $\mathbb{R}$ to ${\cal M}$ defines a
curve [dotted line $\gamma(t)$] on ${\cal M}$, with the points
$\gamma(t_{\rm i})$ and $\gamma(t_{\rm f})$ being the initial
point and the final point of the curve, respectively. The geodesic
mapping $\gamma_0$ corresponds to a curve [solid line
$\gamma_0(t)$] that has the minimal length among curves (in some
neighborhood) that start at a fixed point $\gamma(t_{\rm i})$ and
end at a fixed point $\gamma(t_{\rm f})$. The point $x\in{\cal M}$
is equipped with the tangent space $T_x{\cal M}$ (blue plane),
which contains two vectors $v_x$ and $w_x$. Solid line between the
points $x$ and ${\rm Exp}_x(v_x) \in {\cal M}$ is a geodesic curve
[denoted $\gamma_0^{v_x}(t)$] such that conditions
\eqref{geodesic-v-x} are satisfied. The dotted curve between $x$
and $R_x(v_x)$ is not a geodesic curve in general; however, it has
the same tangent vector at the point $x$ as the geodesic curve
$\gamma_0^{v_x}(t)$. Retraction $R$ is not unique but it defines
alternative point movement $x \rightarrow R_x(v_x)$ and the vector
transport $w_x \rightarrow \tau_{x,v_x}(w_x)$ from the tangent
space $T_x{\cal M}$ to the tangent space $T_{R_x(v_x)}{\cal M}$
(violet plane).}
    \label{fig:R-geometry}
\end{figure}

In this section we informally introduce some necessary preliminaries and notations of the Riemannian geometry. For in-depth introduction the authors recommend Refs.~\cite{absil2009optimization, spivak1970comprehensive, boumal2020introduction}. A manifold ${\cal M}$ of dimension $n$ is a set that can be locally approximated by the Euclidean $n$-dimensional space. A manifold can be seen as a high-dimensional generalization of a smooth surface. For example, the sphere $\{x\in \mathbb{R}^{n+1}| \ \|x\|_2=1\}$ is an $n$-dimensional manifold embedded in $\mathbb{R}^{n+1}$. Each point  $x\in{\cal M}$ is equipped with the tangent space $T_x{\cal M}$, which is an $n$-dimensional vector space. For example, a point $x$ from the two-dimensional unit sphere $\{x\in\mathbb{R}^3|\ \|x\|_2=1\}$ has the tangent space $T_x{\cal M}$ that is merely a tangent plane to the sphere at point $x$, see Fig.~\ref{fig:R-geometry}. One can define a tangent bundle $T{\cal M}$ that is a set consisting of all pairs $(x, T_x{\cal M})$. Each tangent space $T_x{\cal M}$ is equipped with an inner product $g_x:T_x{\cal M}\times T_x{\cal M}\rightarrow \mathbb{R}$. A manifold ${\cal M}$ with the inner product $g_x$ is called a Riemannian manifold. The inner product $g_x$ defines a distance in the neighborhood of a point $x$. A curve on the manifold ${\cal M}$ is a smooth mapping $\gamma:(t_{\rm i}, t_{\rm f})\rightarrow {\cal M}$, where $(t_{\rm i}, t_{\rm f})$ is an interval in $\mathbb{R}$. The length of a curve is defined through the inner product as follows: $L(\gamma) = \int_{t_{\rm i}}^{t_{\rm f}}dt \sqrt{g_{\gamma(t)}\left(\frac{d\gamma(t)}{dt}, \frac{d\gamma(t)}{dt}\right)}$. If $\gamma_0$ is a local minimum of $L(\gamma)$, then $\gamma_0$ is called a geodesic. One can prove that for every $v_x\in T_x{\cal M}$ there exists a unique geodesic $\gamma_0^{v_x}$ such that
\begin{equation} \label{geodesic-v-x}
\gamma_0^{v_x}(0) = x,\quad
\frac{d\gamma_0^{v_x}(t)}{dt}\bigg|_{t=0}=v_x.
\end{equation}

\noindent Using the fact above one can introduce the exponential map \cite{absil2009optimization}.
\begin{definition} \label{exponential_map}
The mapping
\begin{equation}
{\rm Exp}_{x}:T_x{\cal M}\rightarrow {\cal M}, \quad {\rm Exp}_{x}(v_x) = \gamma_0^{v_x}(1)
\end{equation}
is called the exponential map at a point $x$.
\end{definition}

Basic operations in the Euclidean space are the movements of a point (from one position to another) and the vector transport. Definition~\ref{exponential_map} of the exponential map is nothing else but a generalization of the point movement to the case of the Riemannian manifold~\cite{absil2009optimization, lezcano2019cheap, becigneul2018riemannian, li2020efficient}.
However, the exponential map is often computationally inefficient in practice due to a general complexity of geodesics. A less natural but more computationally efficient way to generalize the Euclidean point movement to a Riemannian manifold is to use a retraction defined below~\cite{absil2009optimization}.

\begin{definition} \label{retraction_def}
A retraction on a manifold ${\cal M}$ is a smooth map $R:T{\cal M}\rightarrow {\cal M}$ such that for every $x\in{\cal M}$ the map $R_x: T_x{\cal M}\rightarrow {\cal M}$, $R_x(v_x) \equiv R(x,v_x)$, satisfies
\begin{enumerate}
    \item $R_x(0_x)=x$, where $0_x$ is the zero vector from $T_x{\cal M}$;
    \item $\left. \frac{d}{dt}R_x(tv_x) \right\vert_{t=0}=v_x$ for all $v_x\in T_x{\cal M}$.
\end{enumerate}
\end{definition}

One can see that a set of points $\{R(tv_x)\}_{t \in \mathbb{R}}$
is a smooth curve generalizing the concept of the point movement
beyond the exponential map. If one needs to move a point $x$ along
a curve, whose direction is aligned with the vector $v_x$, then
$R_x(v_x)$ can be considered as a result of that movement, see
Fig.~\ref{fig:R-geometry}. The map $R_x$ is not uniquely defined
in contrast to the exponential map ${\rm Exp}_x$; however, the
former one is usually a computationally cheap alternative to the
latter one. In fact, the exponential map is metric-dependent, so
no analytical expression is known in general (an example is the
manifold of isometric matrices with a non-canonical metric
$g_x$~\cite{edelman1998geometry}). In contrast, the retraction is
metric-independent (see Definition~\ref{retraction_def}), but
$R_x(0_x)={\rm Exp}_x(0_x)$ (it is the point $x$) and
$\frac{d}{dt}R_x(tv_x) \vert_{t=0} = \frac{d}{dt} {\rm Exp}_x(t
v_x) \vert_{t=0}$ (it is the vector $v_x$). Therefore, $R_x(t
v_x)$ is a first-order approximation to ${\rm Exp}_x(t v_x)$ with
respect to a small parameter $t$, see Fig.~\ref{fig:R-geometry}.

The Euclidean vector transport (displacement of a vector along another vector) can be generalized to a Riemannian manifold in the following way~\cite{absil2009optimization, lezcano2019cheap, becigneul2018riemannian, li2020efficient}, see also Fig.~\ref{fig:R-geometry}.
\begin{definition} \label{vector_transport_def}
A vector transport $\tau$ is a smooth map defined on top of a retraction $R$ of a manifold ${\cal M}$ that maps a triple $(x \in {\cal M}, v_x \in T_x {\cal M}, w_x \in T_x {\cal M})$ to a vector $\tau_{x,v_x}(w_x)$
 satisfying the following properties for all $x\in {\cal M}$:
\begin{enumerate}
    \item $\tau_{x,v_x}(w_x)\in T_{R_x(v_x)}{\cal M}$ for all $v_x, w_x\in T_x{\cal M}$ (Underlying retraction);
    \item $\tau_{x,0_x}(w_x)=w_x$ for all $w_x\in T_x{\cal M}$ (Transport along the zero vector is identity);
    \item $\tau_{x,v_x}(aw_x + bu_x) = a\tau_{x,v_x}(w_x) + b\tau_{x,v_x}(u_x)$ for all $a,b \in \mathbb{R}$ and $v_x, w_x, u_x\in T_x{\cal M}$ (Linearity).
\end{enumerate}
\end{definition}

\noindent Similarly to the retraction, the vector transport is not unique. However, the vector transport is usually a computationally cheap alternative to the unique parallel transport, which we do not consider here for the sake of brevity.

Let us consider a basic example of the retraction and the vector transport. Let ${\cal M}$ be a submanifold of $\mathbb{R}^n$, then for a differentiable projection $\pi:\mathbb{R}^n\rightarrow {\cal M}, \ \pi \circ \pi = \pi$, the map
\begin{eqnarray} \label{example-retraction}
R_x(v_x) = \pi(x + v_x)
\end{eqnarray}
is a retraction \cite{lezcano2019cheap}. Let $P_x: \mathbb{R}^n\rightarrow T_x{\cal M}, P_x^2=P_x$ be a linear projector on the tangent space $T_x{\cal M}$. Then the map
\begin{eqnarray} \label{example-vector-transport}
\tau_{x,v_x}(w_x) = P_{R_x(v_x)}(w_x)
\end{eqnarray}
is a vector transport on top of the retraction~\eqref{example-retraction}. One can see that the maps \eqref{example-retraction} and \eqref{example-vector-transport} satisfy Definitions \ref{retraction_def} and \ref{vector_transport_def}, respectively.

\section{Riemannian manifolds in quantum physics}
\label{section-RG-in-QM}


The complex Stiefel manifold is a Riemannain manifold $V_{n, p}$ consisting of all $n\times p$ isometric matrices, $n\geq p$~\cite{edelman1998geometry}. Formally, $V_{n, p}=\{Z\in \mathbb{C}^{n\times p}|Z^\dagger Z=I\}$. If $n=p$, then $V_{n, n}$ is a manifold of unitary $n \times n$ matrices. Unitary matrices describe the time evolution of a closed quantum system as the evolution operator $U=\exp(- i H t)$ is unitary provided the Hamiltonian $H$ is Hermitian. Therefore, the complex Stiefel manifold $V_{n,n}$ is of great use in the circuit quantum computation utilizing unitary gates. Coherent quantum control also operates with unitary transformations~\cite{dong2010quantum,morzhin2019minimal}, so optimization algorithms on the Stiefel manifold of unitary matrices are of great need to efficiently manipulate quantum systems and design quantum algorithms~\cite{garcia2020ibm}.

General $n\times p$ isometric matrices provide a useful parameterization of quantum channels (completely positive and trace preserving maps) via the Stinespring dilation theorem~\cite{holevo2012quantum, stinespring1955positive}. For a given quantum channel, the size of the corresponding isometric matrix depends on the dimension $d$ of the quantum system and the Kraus rank of the channel that does not exceed $d^2$~\cite{holevo2012quantum}. Thanks to the Stinespring dilation, optimization methods on the Stiefel manifold of isometric matrices provide new efficient ways to perform optimization on the set of quantum channels, for instance, to perform the maximum likelihood estimation of quantum channels (process tomography) and  reconstruct both Markovian and non-Markovian open system dynamics~\cite{luchnikov2020machine}. In fact, many algorithms for learning non-Markovian quantum dynamics are based on optimization techniques with restrictions~\cite{luchnikov2020machine, banchi2018modelling, guo2020tensor, krastanov2020unboxing, milz2018reconstructing}.

Another wide area for application of isometric and unitary matrices is quantum tensor networks~\cite{orus2019tensor}.  Some tensor networks such as MERA~\cite{vidal2008class, vidal2009entanglement, evenbly2009algorithms, evenbly2014algorithms} consist of isometric and unitary tensors. As MERA is able to reproduce a wide range of exponentially and polynomially decreasing correlations in multipartite quantum systems, developing an optimization algorithm on the complex Stiefel manifold would allow us to significantly simplify and improve the simulation of many-body quantum systems including the search of low-energy spectrum and eigenstates for local Hamiltonians. In this paper, we accomplish this research programme in Section~\ref{section-ER}. Similar ideas for the use of optimization algorithms on the Stiefel manifold for MERA were independently proposed in Ref.~\cite{hauru-2020}, preprint of which was released a couple of days after the preprint release of the present work~\cite{luchnikov-preprint-2020}.

Another frequently used manifold is the set of density matrices, i.e., Hermitian positive-semidefinite operators with unit trace. Density operators are natural elements of dynamical optimization problems as initial states (say, in the problem of entanglement robustness against a specific noise~\cite{ffk}). The cone of positive-semidefinite matrices comprises both the unnormalized quantum states and the effects of positive operator-valued measures (POVMs), so optimization problems involving denstity operators and effects can be reduced to the optimization problems on the cones of positive-semidefinite operators. Duality between operators and maps on quantum states makes the cone of positive-semidefinite matrices important in the study of general completely positive maps too~\cite{bengtsson2017geometry}. In the present paper, we restrict our analysis to the cone of positive-definite matrices $\mathbb{S}_{++}^n = \{\varrho\in \mathbb{C}^{n\times n}|\varrho=\varrho^\dagger, \ \varrho > 0\}$ and show that such a restriction to positive-definite matrices instead of positive-semidefinite ones does not prohibit approaching boundary points of the cone. We stick to the ground state search by minimizing the variational energy~\cite{umrigar1988optimized, bressanini2002robust} and the experiment-friendly tomography of quantum states by maximizing the likelihood function\cite{d2003quantum, bogdanov2011statistical}. Some other problems where the cone of positive-definite matrices can be of help are listed in Section~\ref{section-concl}.

\section{Riemannian optimization}
\label{section-RO}
\begin{figure}
    \centering
    \includegraphics[width=8cm]{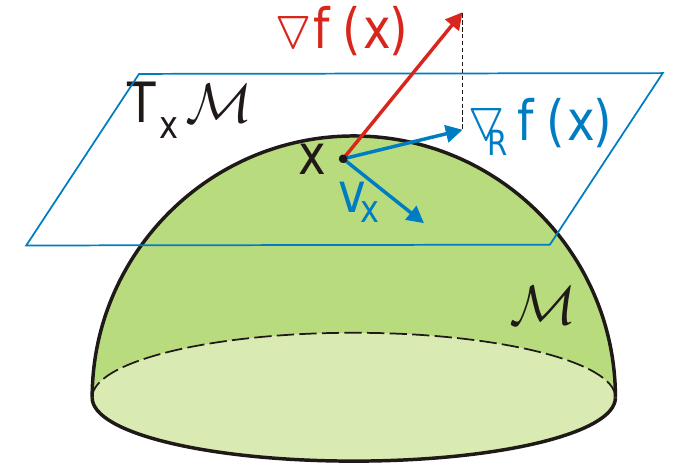}
    \caption{Gradient $\nabla f(x)$ and the Riemannian gradient $\nabla_R f(x)$ at point $x$ of some Riemannian manifold ${\cal M}$. The two gradients are related by Eq.~\eqref{definition-RG}.}
    \label{fig:R-gradient}
\end{figure}

Suppose one wants to find a local minimum of some differentiable function $f(x)$, where $x\in\mathbb{R}^n$. A solution of this optimization problem can be attained by the following iterative scheme called the gradient descent (GD) with momentum~\cite{ruder2016overview}:
\begin{eqnarray}
&& m_{t+1} = \beta m_{t} + (1 - \beta)\nabla f(x_t), \label{momentum-update} \\
&& x_{t+1} = x_t - \eta m_{t+1}, \label{euclid-grad-descent}
\end{eqnarray}

\noindent where $x_t$ is a current approximation of the minimum point, $m_t$ is a vector called momentum, $\eta$ is an optimization step size, $\nabla f(x_t)$ is the gradient of the function $f$ at point $x_t$, and $\beta$ is a constant that ranges from $0$ to $1$ (a typical value is $0.9$). All components of the momentum vector  are initially equal to zero. Starting point $x_0$ is an arbitrary point in $\mathbb{R}^n$. However, if there is a restriction that $x\in{\cal M}$, then the algorithm~\eqref{momentum-update}--\eqref{euclid-grad-descent} fails as it cannot guarantee $x_{t+1} \in{\cal M}$ provided $x_{t} \in{\cal M}$. The algorithm~\eqref{euclid-grad-descent} fails even if $m_{t+1} \in T_{x_t}\mathcal{M}$, so the GD approach has to be modified. In what follows, we consider a generalization of the first-order optimization method~\eqref{momentum-update}--\eqref{euclid-grad-descent}  to Riemannian geometry. To do that we use the concept of the Riemannian gradient at point $x$~\cite{absil2009optimization, lezcano2019cheap, becigneul2018riemannian, li2020efficient} that belongs to a tangent plane $T_x{\cal M}$, see Fig.~\ref{fig:R-gradient}.

To introduce the Riemannian gradient, in addition to the inner product $g_x$ we also need to use some reference inner product $\langle \cdot, \cdot \rangle$ that is independent of $x$. If we dealt with real $n \times p$ matrices $A$ and $B$, then we would use the Frobenius inner product (equivalently, the Hilbert--Schmidt inner product), i.e.,  $\langle A, B \rangle = {\rm tr}(A^T B) = {\rm tr}(A B^T)$~\cite{edelman1998geometry}. For complex matrices $A$ and $B$ we still want the inner product to be a real number, so we follow the lines of Ref.~\cite{sato-2014} and use isomorphic real matrices $\phi(A) = \begin{pmatrix} {\rm Re} \, A & {\rm Im} \, A \\ - {\rm Im} \, A & {\rm Re} \, A \end{pmatrix}$ and $\phi(B) = \begin{pmatrix} {\rm Re} \, B & {\rm Im} \, B \\ - {\rm Im} \, B & {\rm Re} \, B \end{pmatrix}$, respectively, such that $A^{\dag}A = I \Leftrightarrow \phi(A)^T \phi(A) = I$. In view of this, we define the inner product for complex matrices $A$ and $B$ by $\langle A, B \rangle := {\rm tr}[\phi(A)^T \phi(B)] = {\rm tr}(A^{\dag} B + B^{\dag} A) = 2 \, {\rm Re} \, {\rm tr}(A^{\dag} B)$. As the mapping $A \rightarrow \phi(A)$ embeds our manifold of complex matrices in Euclidean space, we refer to the introduced inner product as the Euclidean inner product. The authors of Ref.~\cite{hauru-2020} use a similar definition of the Euclidean inner product (without factor 2).

\begin{definition}\label{riemannian_grad_def}
A vector $\nabla_R f(x)$ is called the Riemannian gradient at point $x\in\mathcal{M}$ if $\nabla_R f(x)\in T_x{\cal M}$ and
\begin{equation} \label{definition-RG}
\left\langle\nabla f(x),v_x\right\rangle = g_x\left(\nabla_Rf(x), v_x\right)\ \text{for all} \ v_x\in T_x{\cal M},
\end{equation}
where $\langle \cdot,\cdot\rangle$ is the Euclidean inner product and $g_x(\cdot,\cdot)$ is the inner product in $T_x{\cal M}$.
\end{definition}

We now proceed to generalization of Eqs.~\eqref{momentum-update}--\eqref{euclid-grad-descent}  to Riemannian geometry. The first step is to replace $\nabla f(x_t)$ to $\nabla_R f(x_t)$:
\begin{equation} \label{m-update}
\tilde{m}_{t+1} = \beta m_{t} + (1 - \beta)\nabla_R f(x_t).
\end{equation}

\noindent The second step is to make a retraction of the modified increment $-\eta \tilde{m}_{t+1}$ at point $x_t$, which results in the new update for the point:
\begin{equation} \label{x-update}
x_{t+1} = R_{x_t} \left(-\eta \tilde{m}_{t+1}\right).
\end{equation}

\noindent If we consider the Riemannian version of an optimization algorithm with the momentum, then one needs to perform the third step and transfer the momentum vector to the new point $x_{t+1}$:
\begin{equation}
m_{t+1} = \tau_{x_t,-\eta \tilde{m}_{t+1}}(\tilde{m}_{t+1}).
\end{equation}

All other first-order optimization methods are generalized to work on the Riemannian manifolds in a similar way. For example, the adaptive moment estimation algorithm ({\scshape Adam}) \cite{kingma2014adam, becigneul2018riemannian, li2020efficient} and the adaptive magnitude stochastic gradient algorithm ({\scshape AMSGrad}) \cite{reddi2019convergence, becigneul2018riemannian} use a modification of Eqs.~\eqref{m-update} and \eqref{x-update}, where three hyperparameters $\beta_1$, $\beta_2$, and $\epsilon$ are used instead of the single hyperparameter $\beta$. In the present work, we consider generalized versions of the conventional GD, the GD with momentum, {\scshape Adam}, and {\scshape AMSGrad}.

\section{Automatic differentiation}
\label{section-AD-and-RO}

One of the most practical ways of using the Riemannian optimization is to combine it with the automatic differentiation algorithm~\cite{baydin2017automatic}. The automatic differentiation redirects the whole procedure of the gradient calculation to a computer. It dramatically simplifies the implementation of various numerical algorithms. The automatic differentiation is essentially an algorithmic way of thinking about the chain rule for derivatives. Assume one needs to evaluate a derivative of a composite function ${\cal L}:\mathbb{R}^n\times \mathbb{R}^m\rightarrow \mathbb{R}^k$,
\begin{equation*}
{\cal L}(x, y) = f \Big( g \big( q(x), p(y) \big) \Big),
\end{equation*}

\noindent at a point $(x_0, y_0)$. Here, $f: \mathbb{R}^l \to \mathbb{R}^k$, $g: \mathbb{R}^s \times \mathbb{R}^t \to \mathbb{R}^l$, $q: \mathbb{R}^n \to \mathbb{R}^s$, and $p: \mathbb{R}^m \to \mathbb{R}^t$ are elementary functions with predefined derivatives, i.e., a computer can evaluate a $k \times l$ matrix $\frac{\partial f(g)}{\partial g}\big|_{g=g_0}$, an $l \times s$ matrix $\frac{\partial g(q, p_0)}{\partial q}\big|_{q=q_0}$, an $l \times t$ matrix $\frac{\partial g(q_0, p)}{\partial p}\big|_{p=p_0}$, an $s \times n$ matrix $\frac{\partial q(x)}{\partial x}\big|_{x=x_0}$, and a $t \times m$ matrix $\frac{\partial p(y)}{\partial y}\big|_{y=y_0}$ with the machine precision. The automatic differentiation algorithm consists of two steps. At the first step (called the forward propagation) we sequentially evaluate the values of all functions at the given point $(x_0, y_0)$:
\begin{equation*}
q_0 = q(x_0),\ p_0 = p(y_0) \quad \rightarrow \quad g_0 = g(q_0, p_0) \quad \rightarrow \quad f_0 = f(g_0).
\end{equation*}

\noindent At the second step we sequentially evaluate derivatives of a function ${\cal L}$ with respect to all intermediate variables. For the sake of simplicity we denote $\frac{\partial {\cal L}}{\partial x} = \bar{x}$. In this notation, $\bar{{\cal L}} = \frac{\partial {\cal L}}{\partial{\cal L}} = I$, where $I$ is the $k \times k$ identity matrix. Following the chain rule, we sequentially get
\begin{equation*}
\bar{g} = \bar{{\cal L}} \left.\frac{\partial {\cal L}(g)}{\partial g}\right\vert_{g=g_0} \quad \rightarrow \quad \bar{q}=\bar{g} \left.\frac{\partial g(q, p_0)}{\partial q}\right\vert_{q=q_0}, \ \bar{p}=\bar{g} \left. \frac{\partial g(q_0, p)}{\partial p} \right\vert_{p=p_0} \quad \rightarrow \quad \bar{x}=\bar{q} \left.\frac{\partial q(x)}{\partial x}\right\vert_{x=x_0}, \ \bar{y}=\bar{p} \left. \frac{\partial p(y)}{\partial y} \right\vert_{y=y_0}.
\end{equation*}

\noindent This step is called the back propagation because it progresses in the opposite direction in comparison with the forward propagation. Combining the forward and backward propagations, we evaluate the desired derivatives of ${\cal L}$ with the machine precision. This procedure is automatically implemented for arbitrary composite functions, e.g., in TensorFlow~\cite{tensorflow}, and is known as the automatic differentiation. The automatic differentiation is used for the gradient-based training of various machine learning models including neural networks and tensor networks~\cite{zhang2019automatic}. In the following sections we show that the automatic differentiation is applicable to the calculation of gradients of objective functions during the Riemannian optimization. It makes the application of the Riemannian optimization easy and straightforward to any problem, e.g., the variational energy minimization for tensor networks such as MERA.

\section{Riemannian optimization on the Stiefel manifold}
\label{section-Stiefel}

Consider the Stiefel manifold $V_{n,p}$ of $n \times p$ isometric matrices with complex entries, $n \geq p$. Suppose $X \in V_{n,p}$, then $\exp(-iHt) X \in V_{n,p}$ if $t\in \mathbb{R}$ and $H$ is an $n \times n$ Hermitian matrix. The tangent space $T_X V_{n,p}$ consists of all $n \times p$ matrices of the form $-i H X$, $H=H^{\dag}$. By definition of the Riemannian manifold, we must introduce the inner product $g_X$ for elements of the tangent space $T_X V_{n,p}$. If $g_X(A,B) = 2  {\rm Re} \, {\rm tr}(A^{\dag}B)$, then this inner product is independent of $X$ and defines the Euclidean metric~\cite{sato-2014}. The so-called canonical metric on the Stiefel manifold is a bit different and reads $g_X(A,B) = 2 {\rm Re} \, {\rm tr}[A^{\dag} (I - \frac{1}{2}XX^{\dag}) B]$, Ref.~\cite{edelman1998geometry}. Consider a differentiable function $f: V_{n,p} \rightarrow \mathbb{R}$ and its conventional gradient $\nabla f(X)$, which we treat as an $n \times p$ matrix with elements $\partial f / \partial X_{ij}$, $i=1,\ldots,n$, $j=1,\ldots,p$. In what follows, we derive the Riemannian gradient $\nabla_R f(X)$ for two types of metric.

1. The Euclidean metric $g_X(A,B) = 2  {\rm Re} \, {\rm tr}(A^{\dag}B)$. Eq.~\eqref{definition-RG} reduces to
\begin{equation*}
{\rm Re} \, {\rm tr} \left[ \Big( \nabla f(X) - \nabla_R f(X) \Big)^{\dag} (-iHX) \right] = 0,
\end{equation*}

\noindent which is to be valid for all Hermitian matrices $H$ (i.e., for all tangent vectors $-iHX$). This implies Hermiticity of the matrix $X \Big( \nabla f(X) - \nabla_R f(X) \Big)^{\dag}$, i.e.,
\begin{equation*}
X \Big( \nabla f(X) - \nabla_R f(X) \Big)^{\dag} = \Big( \nabla f(X) - \nabla_R f(X) \Big) X^{\dag}.
\end{equation*}

\noindent Multiplying both sides of the obtained equation by $X$ and taking into account the relation $X^{\dag}X = I$ (the $p \times p$ identity matrix), we have
\begin{equation} \label{derivation-RG-Hermiticity-condition}
X \big(\nabla f(X) \big)^{\dag} X - X \big( \nabla_R f(X) \big)^{\dag} X = \nabla f(X) - \nabla_R f(X).
\end{equation}

\noindent Since $\nabla_R f(X)$ belongs to the tangent space $T_X V_{n,p}$ by definition of the Riemannian gradient, we have $\nabla_R f(X) = - i H' X$ for some Hermitian matrix $H'$ and, therefore, $(\nabla_R f(X))^{\dag} X = i X^{\dag} H' X = - X^{\dag} \nabla_R f(X)$. Substituting this expression in Eq.~\eqref{derivation-RG-Hermiticity-condition}, we get
\begin{equation*}
(I + X X^{\dag}) \nabla_R f(X) = \nabla f(X) -  X \big(\nabla f(X) \big)^{\dag} X,
\end{equation*}

\noindent where $I$ is the $n \times n$ identity matrix. Note that
$XX^{\dag}$ is a rank-$p$ projector, so the operator $I + X
X^{\dag}$ is readily inverted, namely, $(I + X X^{\dag})^{-1} = I
- \frac{1}{2} X X^{\dag}$. Hence,
\begin{eqnarray}
 \nabla_R f(X) &=& (I - \frac{1}{2} X X^{\dag}) \left[ \nabla f(X) -  X \big(\nabla f(X) \big)^{\dag} X \right] \nonumber\\
 &=& \nabla f(X) - \frac{1}{2} X \Big[ X^\dagger \nabla f(X) + \big(\nabla f(X)\big)^\dagger X \Big]. \label{RG-Stiefel-Euclidean}
\end{eqnarray}

\noindent If $X$ is unitary ($n=p$), then $\nabla_R f(X) = \frac{1}{2} \left[ \nabla f(X) -  X \big(\nabla f(X) \big)^{\dag} X \right]$.

2. The canonical metric $g_X(A,B) = 2 {\rm Re} \, {\rm tr}[A^{\dag} (I - \frac{1}{2}XX^{\dag}) B]$. Eq.~\eqref{definition-RG} reduces to
\begin{equation*}
{\rm Re} \, {\rm tr} \left[ \Big( \nabla f(X) - (I - \frac{1}{2}XX^{\dag}) \nabla_R f(X) \Big)^{\dag} (-iHX) \right] = 0,
\end{equation*}

\noindent which is to be valid for all Hermitian matrices $H$ (i.e., for all tangent vectors $-iHX$). This implies Hermiticity of the matrix $X \Big( \nabla f(X) - (I - \frac{1}{2}XX^{\dag}) \nabla_R f(X) \Big)^{\dag}$, i.e.,
\begin{equation*}
X \Big( \nabla f(X) - (I - \frac{1}{2}XX^{\dag}) \nabla_R f(X) \Big)^{\dag} = \Big( \nabla f(X) -  (I - \frac{1}{2}XX^{\dag}) \nabla_R f(X) \Big) X^{\dag}.
\end{equation*}

\noindent Multiplying both sides of the obtained equation by $X$ and taking into account the relation $X^{\dag}X = I$ (the $p \times p$ identity matrix), we have
\begin{equation} \label{derivation-RG-Hermiticity-condition-2}
X \big(\nabla f(X) \big)^{\dag} X - \frac{1}{2} X \big( \nabla_R f(X) \big)^{\dag} X = \nabla f(X) - (I - \frac{1}{2}XX^{\dag}) \nabla_R f(X).
\end{equation}

\noindent Since $\nabla_R f(X)$ belongs to the tangent space $T_X V_{n,p}$ by definition of the Riemannian gradient, we have $\nabla_R f(X) = - i H' X$ for some Hermitian matrix $H'$ and, therefore, $(\nabla_R f(X))^{\dag} X = i X^{\dag} H' X = - X^{\dag} \nabla_R f(X)$. Substituting this expression in Eq.~\eqref{derivation-RG-Hermiticity-condition-2}, we get
\begin{equation} \label{RG-Stiefel-canonical}
\nabla_R f(X) = \nabla f(X) -  X \big(\nabla f(X) \big)^{\dag} X.
\end{equation}

The next step in the Riemannian optimization is to move from one point $X$ to another point $R_X (W)$ along a curve, whose direction in the first point (i.e., vector $W$) is defined by the Riemannian gradient. As there exist various retractions $R_X(W)$ for the Stiefel manifold $V_{n,p}$, we focus on two of them.

1. The Cayley retraction reads~\cite{li2020efficient, nishimori2005learning}
\begin{eqnarray} \label{Cayley-retraction}
&& R_X^{\rm Cayley}(W) = \left( I + \frac{i H}{2} \right)^{-1} \left( I - \frac{i H}{2} \right) X, \\
&& H = H^{\dag} = i(WX^{\dag}-XW^{\dag}) - \frac{i}{2}\left[ X (X^{\dag}WX^{\dag}) - (XW^{\dag}X)X^{\dag} \right]. \nonumber
\end{eqnarray}

\noindent A simple vector transport based on the Cayley retraction is given by Eq.~\eqref{example-vector-transport}, where
\begin{equation} \label{projector-onto-tangent-Stiefel}
P_X(Y) = Y - \frac{1}{2} X \left( Y^{\dag} X + X^{\dag} Y \right)
\end{equation}

\noindent is a projector because $P_X \big( P_X(Y) \big) = P_X(Y)$ in view of $X^{\dag}X = I$.

2. The singular value decomposition (SVD) $A = U \Sigma V^{\dag}$ of an $n \times p$ matrix $A$ enables one to construct a projection $\pi$ onto the manifold of $n \times p$ isometric matrices through the relation $\pi(A) = UV^{\dag}$. The induced SVD retraction reads~\cite{absil2012projection}
\begin{equation} \label{SVD-retraction}
    R_X^{\rm SVD}(W) = \pi(X + W),
\end{equation}

\noindent and the vector transport is given by Eq.~\eqref{example-vector-transport}, with the projector on the tangent space being \eqref{projector-onto-tangent-Stiefel}.

A Riemannian optimization algorithm for minimization of a differentiable function $f(X)$ on the Stiefel manifold ($X \in V_{n,p}$) with a specified metric is constructed by combining the Riemannian gradient $\nabla_R f(X)$, the retraction $R_X(W)$, and the vector transport $\tau_{X,W}(Z)$. As an illustrative example we consider the Stiefel manifold with the Euclidean metric and provide a step-by-step algorithm for the gradient descent (GD) with momentum. Depending on the chosen retraction type (the Cayley retraction or the SVD retraction), there appears a difference in one step only. The algorithm reads as follows:
\begin{enumerate}
    \item Choose an initial isometric matrix $X_0$ at random.
    \item Set the step size $\eta$ and the value of the hyperparameter $\beta$.
    \item \label{loop-start} Compute the gradient of the objective function $\nabla f(X_t)$ at a current point $X_t$ of the manifold ($t=0$ in the beginning).
    \item Compute the Riemannian gradient in the Euclidean metric via formula~\eqref{RG-Stiefel-Euclidean}, i.e., $\nabla_R f(X_t) = \nabla f(X_t) - \frac{1}{2} X_t \Big[ X_t^\dagger \nabla f(X_t) + \big(\nabla f(X_t)\big)^\dagger X_t \Big]$.
    \item Compute the update direction $\tilde{m}_{t+1} = \beta m_t + (1 - \beta) \nabla_R f(X_t)$, where $m_0 = 0$.
    \item Move to another point of the manifold by using the formula $X_{t+1} = R^{\rm Cayley}_{X_t}(-\eta \tilde{m}_{t+1})$ if the Cayley retraction is chosen, see Eq.~\eqref{Cayley-retraction}, or the formula $X_{t+1} = R_{X_t}^{\rm SVD}(-\eta \tilde{m}_{t+1})$ if the SVD retraction is chosen, see Eq.~\eqref{SVD-retraction}.
    \item \label{loop-end} Implement the vector transport of the update direction $\tilde{m}_{t+1}$ to the new point $X_{t+1}$ via the orthogonal projection on the tangent space of $X_{t+1}$, i.e., calculate $m_{t+1} = P_{X_{t+1}}(\tilde{m}_{t+1})$, see Eq.~\eqref{projector-onto-tangent-Stiefel}.
    \item Repeat steps \ref{loop-start} to \ref{loop-end} for a desired number of iterations.
\end{enumerate}

\noindent If the canonical metric is used, the expression for the Riemannian gradient in step 4 is to be modified accordingly, see Eq.~\eqref{RG-Stiefel-canonical}.

It is also instructive to discuss complexity of the presented algorithm. The complexity of step 3 entirely depends on the structure of $f$. Since we use the automatic differentiation to evaluate the gradient of $f$, the complexity of this step is the same as the complexity of the function evaluation. Therefore, the complexity of step 3 is task-specific and we do not take it into account in what follows; however, as the gradient is an $n \times p$ matrix too, we conclude that the complexity of the gradient calculation is greater than or equal to $O(np)$. The complexity of step 4 is $O(np^2)$, see Ref.~\cite{luchnikov-2021}. Step 5 is elementary and requires $O(np)$ summation and multiplication operations. The complexity of step 6 is $O(n^3)$ in the case of the Cayley retraction and $O(np^2)$ in case of the SVD retraction, see Ref.~\cite{luchnikov-2021}. The complexity of step 7 is $O(np^2)$, see Ref.~\cite{luchnikov-2021}. As result, the overall complexity is $O(n^3)$ in the case of the Cayley retraction and $O(np^2)$ in the case of the SVD retraction.

In the following subsections we implement the Riemannian optimization on the Stiefel manifold to solve some particular problems of quantum physics and analyze its performance. In Section~\ref{section-Hamiltonian-renorm}, we consider the Hamiltonian renormalization to study the low-energy properties. In subsequent Sections~\ref{section-entangling}, \ref{section-state-preparation}, and \ref{section-design}, we consider certain problems in the very diverse field of optimal quantum control~\cite{dong2010quantum, plesch2011quantum, altafini2012modeling}. For each problem, we calculate the cost function gradient by using the automatic differentiation.

\subsection{Low-energy spectrum and eigenstates} \label{section-Hamiltonian-renorm}

We begin with a problem of the Hamiltonian renormalization, which aims at replacing a high-dimensional Hamiltonian $H$ by a simpler low-dimensional one with the identical low-energy spectrum. We consider this problem in order to compare performances of the conventional MERA optimizer~\cite{evenbly2009algorithms} and the first-order Riemannian optimizers. We show that the first-order Riemannian optimizers with properly chosen hyperparameters outperform the conventional MERA optimizer. Suppose $H$ is an $n \times n$ matrix, then the renormalization transform has the form $H \rightarrow H_{\rm r} = V_{0}^\dagger H V_{0}$, where $V_{0}$ is an $n \times p$ isometric matrix such that $V_0^{\dag}$ maps $p$ eigenstates of $H$ with the lowerest energy to the eigenstates of $H_{\rm r}$, $p<n$. As $n$ is assumed to be large, we do not have access to the spectrum and eigenstates of $H$. Instead, the linear functional $L(V) := {\rm tr} \left(V^\dagger H V\right)$ is readily computable, so we solve the optimization problem
\begin{equation} \label{min-L(V)}
\begin{aligned}
   &\Min \quad L(V) \\
   &\text{subject to}\quad V\in V_{n, p}. \\
\end{aligned}
\end{equation}

\noindent Clearly, $\min_{V\in V_{n, p}} L(V) = L(V_0)$. On the other hand, $L(V_0) = \sum_{i=1}^p \lambda_i$, where $\{\lambda_i\}_{i=1}^n$ are the eigenvalues of the Hamiltonian $H$ in the ascending order. The low-energy eigenvectors $\ket{\psi_k}$ of $H$ are expressed through the eigenvectors $\ket{\varphi_k}$ of $H_{\rm r}$ via $\ket{\psi_k} = V_0\ket{\varphi_k}$, $k=1,\ldots,p$. The exact minimal value of $L$ (if known) can be used to evaluate the performance of the optimization algorithm.

\begin{figure}
    \centering
    \includegraphics[width=12cm]{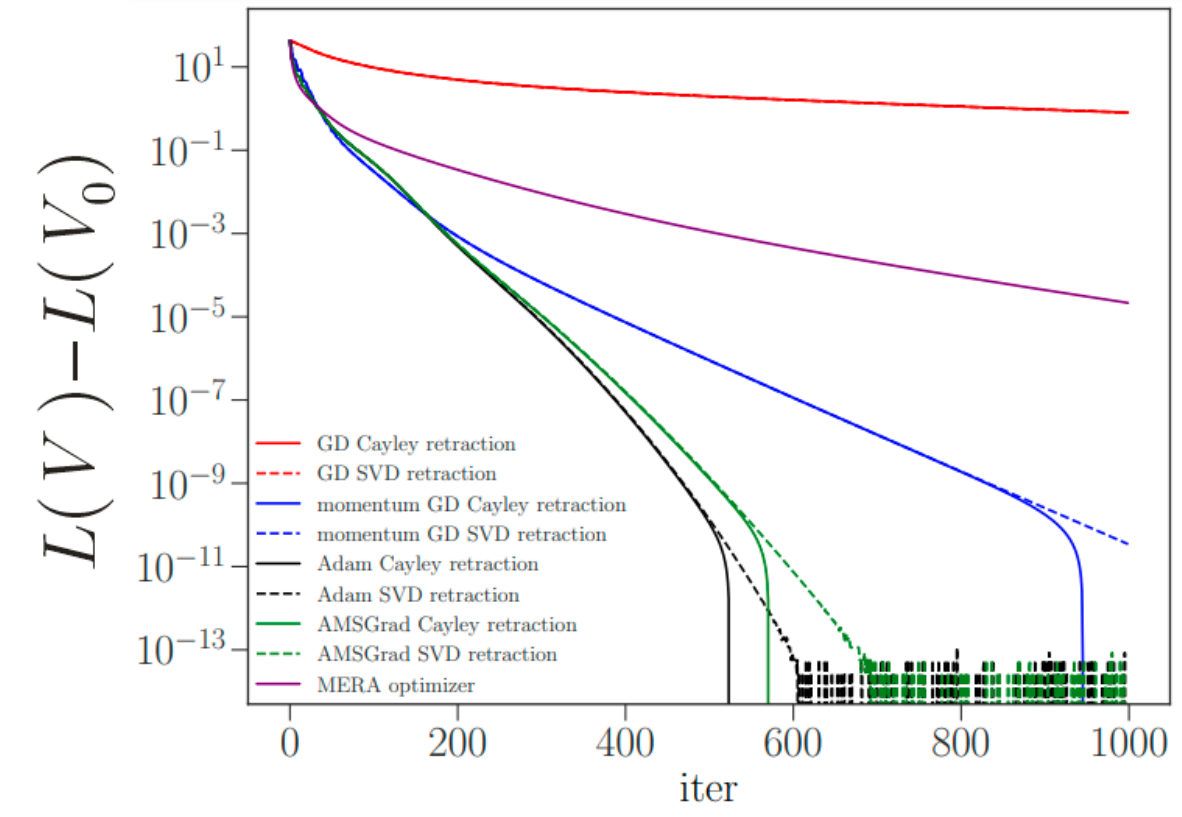}
    \caption{Optimization performance in search of the low-energy spectrum \eqref{min-L(V)} for various Riemannian optimization algorithms: Difference $L(V)-L(V_0)$ between the value $L(V)$ at a given iteration of the algorithm and the exact minimum $L(V_0)$ vs iteration number. The Hamiltonian dimension is $n=100$, the number of low-energy levels is $p=30$. The step size $\eta$ for the Riemannian GD, the Riemannian GD with momentum, the Riemannian {\scshape Adam} and the Riemannian {\scshape AMSGradand} is $0.003$ (the highest possible one guaranteeing convergence), $0.2$, $0.3$, and $0.3$, respectively. Other hyperparameters are standard for the algorithms ($\beta = 0.9$ for the GD with momentum; $\beta_1 = 0.9$, $\beta_2 = 0.999$, $\epsilon=10^{-8}$ for {\scshape Adam} and {\scshape AMSGradand}).}
    \label{methods_comparison_stiefel}
\end{figure}

Results of the Riemannian optimization with the Euclidean metric are presented in Fig.~\ref{methods_comparison_stiefel} for various algorithms exploiting either the Cayley retraction or the SVD retraction: the Riemannian GD, the Riemannian GD with momentum, the Riemannian {\scshape Adam}, the Riemannian {\scshape AMSGrad}, and the conventional MERA optimizer. Here, we put $n=100$ and $p=30$ to monitor the optimization perfrormance. To make the optimization challenging, we generate a random ill-conditioned Hamiltonian matrix $H$ as follows: (i) we generate samples $\{s_i\}_{i=1}^n$ from the uniform distribution on the segment $[-4, 0]$; (ii) then we exponentiate samples and subtract the maximum value out of each exponent: $\lambda_i = \exp(s_i) - \max_{i}\exp(s_i)$; (iii) we generate a random unitary matrix $U$ using QR decomposition and build a Hamiltonian $H = U^\dagger\Lambda U$, where $\Lambda = {\rm diag}(\{\lambda_i\}_{i=1}^n)$. The ill-conditioned nature of spectrum makes the optimization problem difficult to solve by the naive Riemannian GD because it forbids to pick a reasonably large step size. Negativity of spectrum allows us to compare the performance of the Riemannian optimization technique with the conventional MERA optimization algorithm that requires the Hamiltonian to be negatively defined~\cite{evenbly2009algorithms} (see Section~\ref{MERA_optimizer} for details on the conventional MERA optimizer). Fig.~\ref{methods_comparison_stiefel} demonstrates that the conventional MERA optimizer performs worse than the Riemannian {\scshape Adam}, the Riemannian {\scshape AMSGrad}, and the Riemannian GD with momentum.

\subsection{Preparation of highly entangled states} \label{section-entangling}

\begin{figure}
    \centering
\includegraphics[width=15cm]{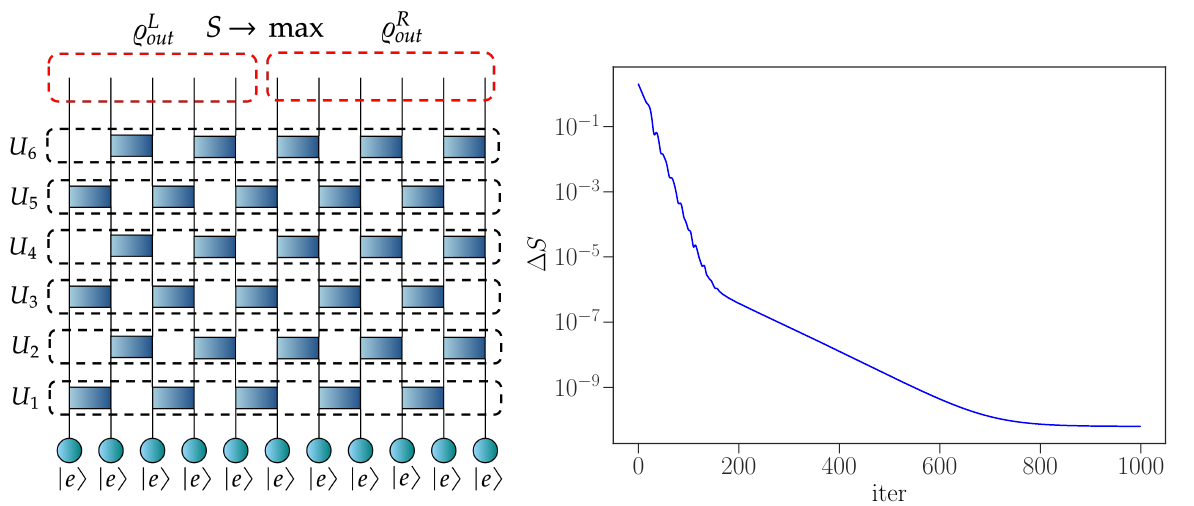}
    \caption{Left: Tensor diagram of unitary two-qubit gates (boxes) governing initially disentangled qubits toward the  maximally entangled state (with respect to the red dotted cut between 5 and 6 qubits). Each unitary gate is tuned by the Riemannian optimization algorithm. The gradient is calculated via the automatic differentiation. Right: Performance of the Riemannian {\scshape AMSGrad} with the Euclidean metric on the complex Stiefel manifold and the SVD retraction in achieving the maximally entangled state. Number of unitary layers $M=6$, the algorithm step size $\eta=0.1$, other hyperparameters are standard ($\beta_1 = 0.9$, $\beta_2 = 0.999$, $\epsilon=10^{-8}$). Difference $\Delta S = S_2^{\rm max} - S_2$ quantifies the deviation of the prepared state from the maximally entangled one at a given iteration of the algorithm, see Eq.~\eqref{ent-purity-min}.}
    \label{entropy_control_tn}
\end{figure}

Suppose one can experimentally implement any two-qubit unitary
transformation on adjacent qubits arranged in line
(Fig.~\ref{entropy_control_tn}). The problem is to find such
unitary transformations that the final state exhibits the maximum
entanglement with respect to a given cut provided all the qubits
are initially disentangled, $\ket{\psi_{\rm in}}=\ket{e}\otimes
\ket{e}\otimes\dots\otimes\ket{e}$. We consider a cut separating
$N^L$ qubits on the left from $N^R$ qubits on the right. Combining
commuting unitary gates into multiqubit operators $U_{2m+1} =
u_1^{(m)} \otimes u_2^{(m)} \otimes \ldots \otimes I$ and $U_{2m}
= I \otimes v_1^{(m)} \otimes v_2^{(m)} \otimes \ldots$ (depicted
by dashed layers in Fig.~\ref{entropy_control_tn}), the final
state is $\ket{\psi_{\rm out}} = U_M \dots U_2 U_1 \ket{\psi_{\rm
in}}$ for some $M$ determining the network depth. As the pure
state entanglement can be quantified via the spectrum of the
reduced density operator ($\varrho_{\rm out}^{L}$ or $\varrho_{\rm
out}^{R}$), we end up with the following optimization problem:

\begin{equation} \label{ent-purity-min}
\begin{aligned}
   &\Max \quad S_2 := - \log\left( {\rm tr}\left[(\varrho_{\rm out}^L)^2\right] \right) = - \log\left({\rm tr}\left[(\varrho_{\rm out}^R)^2 \right] \right) \\
   &\text{subject to}\quad u_j^{(m)}, v_j^{(m)} \in V_{4, 4}\quad \text{~for~all~}j,m. \\
\end{aligned}
\end{equation}

\noindent Here, $S_2$ is the second order R\'{e}nyi entropy. The theoretical maximum in Eq.~\eqref{ent-purity-min} equals $S_2^{\rm max} = \log \min(N^L, N^R)$ and corresponds to the maximally entangled state, which can be realized in a sufficiently deep network.

We solve the optimization problem \eqref{ent-purity-min} for $N^L = 5$ and $N^R = 6$ on the Stiefel manifold of $4\times 4$ unitary matrices by using the Riemannian {\scshape AMSGrad} algorithm with the step size $\eta = 0.1$. We show that $M \leq 5$ layers in Fig.~\ref{entropy_control_tn} are not enough to approach the theoretical maximum and prepare the maximally entangled state. $M \geq 6$ layers suffice to achieve the desired maximally entangled state and the Riemannian {\scshape AMSGrad} successfully solves the problem. To address the optimization process performance with $M = 6$ layers we plot the deviation $\Delta S = S_2^{\rm max} - S_2$ as a function of the number of iterations in the Riemannian {\scshape AMSGrad} (Fig.~\ref{entropy_control_tn}).

\subsection{Optimal control for state preparation} \label{section-state-preparation}

\begin{figure}
    \centering
    \includegraphics[width=15cm]{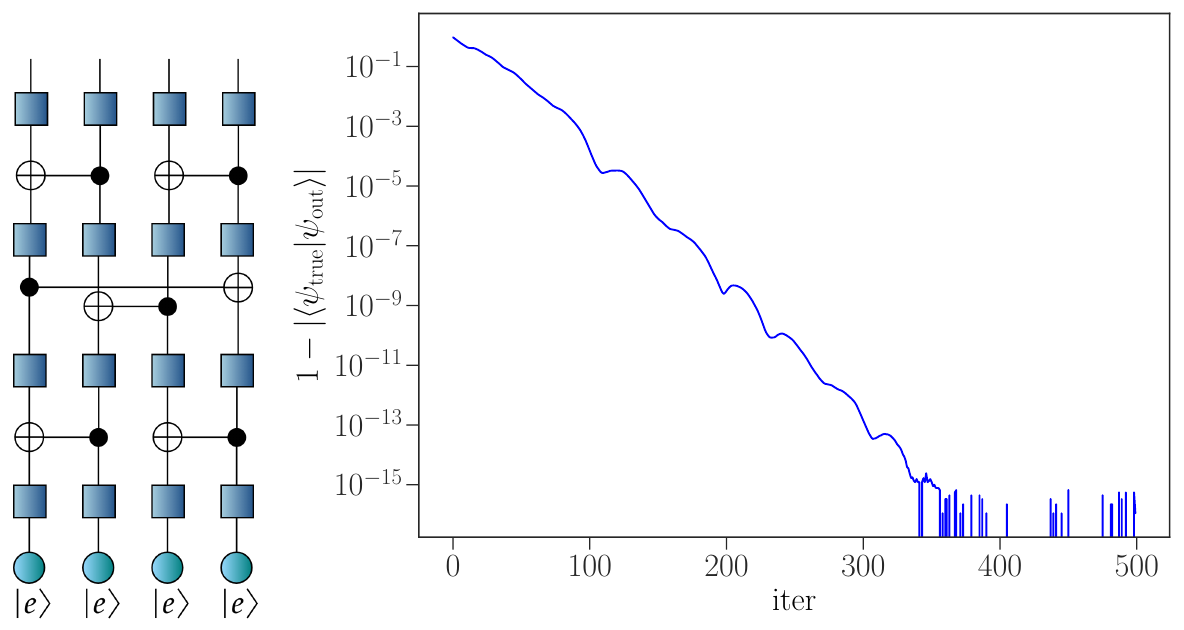}
    \caption{Left: Quantum circuit with {\scshape CNOT} gates and local unitary gates to be adjusted in a such way that the circuit outputs a desired 4-qubit state. Right: Performance of the Riemannian {\scshape Adam} algorithm with the Euclidean metric on the complex Stiefel manifold and the SVD retraction in solving the state preparation problem \eqref{state-preparation-opt-problem}. Step size $\eta =0.05$, other hyperparameters are standard ($\beta_1 = 0.9$, $\beta_2 = 0.999$, $\epsilon=10^{-8}$).}
    \label{state_prep_tn}
\end{figure}

The problem is to prepare a desired quantum state of several qubits by using a quantum circuit, whose architecture allows using {\scshape CNOT} gates and local (one-qubit) unitary gates as it takes place, e.g., in IBM quantum processors~\cite{ibm}. The initial state of the qubits is factorized, i.e., $\ket{\psi_{\rm in}} = \ket{e}^{\otimes N}$, where $N$ is a number of qubits. Suppose the architecture of {\scshape CNOT} gates is fixed but one can apply arbitrary local unitary gates, then the output state is $\ket{\psi_{\rm out}} = \prod_{j=1}^M \left( U_j \bigotimes_{k=1}^N u_{jk} \right) \ket{\psi_{\rm in}}$, where $\{u_{jk}\}_{j=1, \ldots, M; \, k=1,\ldots,N}$ is a set of one-qubit unitary gates, $M$ is a number of unitary layers alternating with layers of {\scshape CNOT} gates, $U_j = \prod_{i} \text{\scshape CNOT}_{p_i q_i}^{(j)}$ if $j<M$, $U_M = I$, and $\text{\scshape CNOT}_{p_i q_i}^{(j)}$ is a {\scshape CNOT} gate with the controlling qubit $p_i$ and the controlled qubit $q_i$ within the $j$-th layer, see Fig.~\ref{state_prep_tn}. The goal is to make the output state $\ket{\psi_{\rm out}}$ as close to a desirable state $\ket{\psi_{\rm true}}$ as it is possible. This leads to the following optimization problem:
\begin{equation} \label{state-preparation-opt-problem}
\begin{aligned}
   &\Max \quad |\langle \psi_{\rm true}|\psi_{\rm out}\rangle| \\
   &\text{subject to}\quad u_{jk}\in V_{2, 2}\quad  \text{~for~all~}j,k. \\
\end{aligned}
\end{equation}

We solve this problem by using the Riemannian optimization on the Stiefel manifold of unitary matrices and the automatic differentiation. As an illustrative example we consider $N=4$ and $M=4$ for a circuit depicted in Fig.~\ref{state_prep_tn} and a randomly chosen state $\ket{\psi_{\rm true}}$. The right-hand side of Fig.~\ref{state_prep_tn} demonstrates that the Riemannian {\scshape Adam} optimizer with the step size $\eta = 0.05$ rapidly finds such unitary gates that the overlap $|\langle \psi_{\rm true}|\psi_{\rm out}\rangle|$ equals 1 up to the machine precision.

\begin{figure}
    \centering
    \includegraphics[width=12cm]{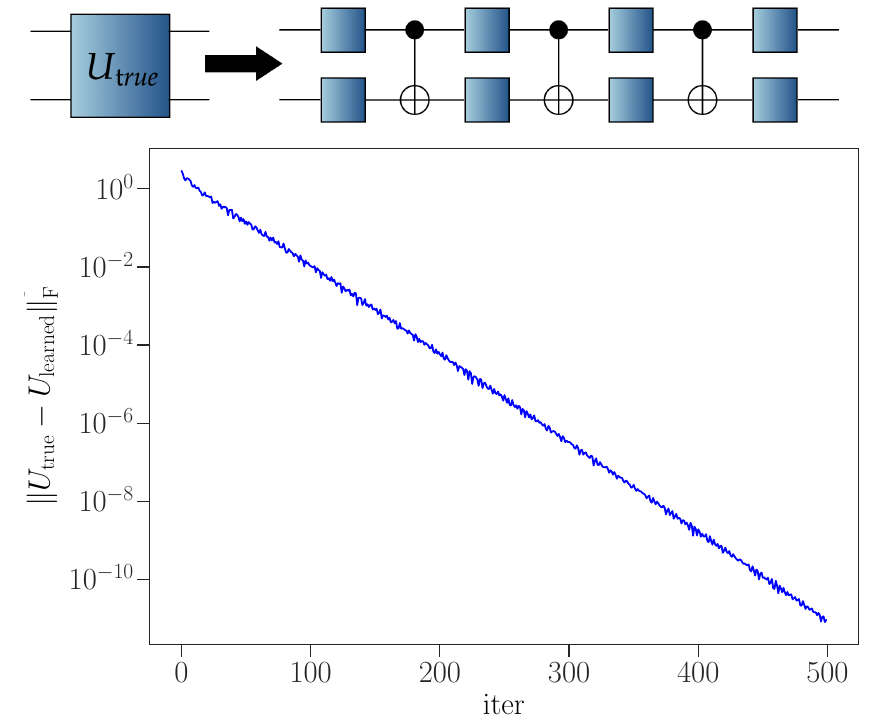}
    \caption{Top: Two-qubit gate decomposition problem. Bottom: Performance of the Riemannian {\scshape AMSGrad} optimization algorithm with the Euclidean metric on the complex Stiefel manifold and the SVD retraction in solving the decomposition problem~\eqref{frobenius-distance-unitary}. Step size $\eta =0.2$, other hyperparameters are standard ($\beta_1 = 0.9$, $\beta_2 = 0.999$, $\epsilon=10^{-8}$).}
    \label{gate_decomposition_tn}
\end{figure}

\subsection{Design and decomposition of quantum gates} \label{section-design}

Decomposition of a desired multiqubit unitary gate $U_{\rm true}$
into simpler constituents is of vital importance for circuit
implementation of quantum computing~\cite{ibm}. Elementary gates
usually include the entangling {\scshape CNOT} gate and one-qubit
gates. The general aim is to find such a decomposition of a
multiqubit unitary gate that involves the least possible number of
elementary gates (or {\scshape CNOT} gates as they are usually
much noisier than the one-qubit ones). This problem is addressed
by minimizing of the square of Frobenius distance between the
desired gate $U_{\rm true}$ and its current decomposition $U_{\rm
learned} = \prod_{j=1}^M \left( U_{j} \bigotimes_{k=1}^N u_{jk}
\right)$ as in Section~\ref{section-state-preparation}. If the
distance vanishes, then the decomposition is valid. However, there
is a need of discrete optimization over architectures of {\scshape
CNOT} gates in general. Here we consider a simpler problem wherein
the architecture of {\scshape CNOT} gates is fixed but one-qubit
unitary gates $u_{jk}$ are optimized. This leads to the following
optimization problem:

\begin{equation} \label{frobenius-distance-unitary}
\begin{aligned}
   &\Min \quad \|U_{\rm true} - U_{\rm learned}\|_{F}^2 \\
   &\text{subject to}\quad u_{jk} \in V_{2, 2}\quad \text{~for~all~}j,k. \\
\end{aligned}
\end{equation}

To provide a proof-of-principle solution of problem
\eqref{frobenius-distance-unitary} with the use of the Riemannian
optimization on the Stiefel manifold of unitary matrices, we
consider a randomly generated two-qubit gate $U_{\rm true}$ and
decompose it into the tensor network with three {\scshape CNOT}
gates depicted in Fig.~\ref{gate_decomposition_tn}. The minimum in
\eqref{frobenius-distance-unitary} is known to be zero in this
case~\cite{kraus-2001}. Bottom part of
Fig.~\ref{gate_decomposition_tn} clearly shows that the Riemannian
{\scshape AMSGrad} optimization with the step size $\eta=0.2$
successfully solves the posed problem and finds the optimal gates
$\{ u_{jk} \}$, for which the distance
\eqref{frobenius-distance-unitary} becomes negligible.

\section{Riemannian optimization for the entanglement renormalization}
\label{section-ER}

\subsection{Tensor network describing the multiscale entanglement-renormalization ansatz}

\begin{figure}
    \centering
    \includegraphics[width=15cm]{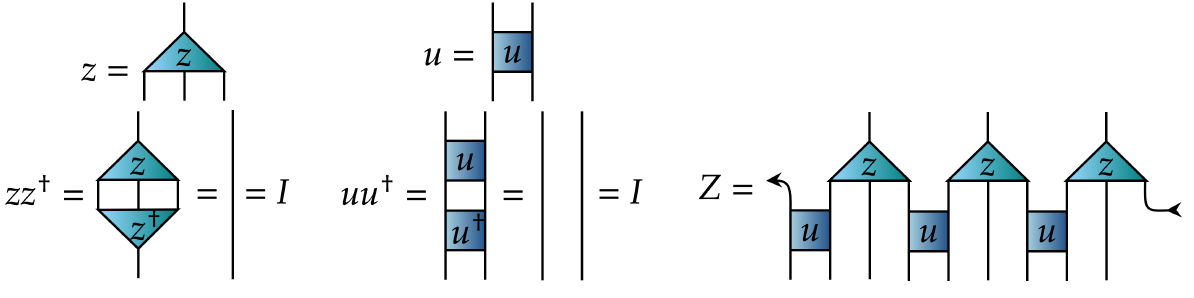}
    \caption{Left to right: Diagrammatic representation of the elementary isometry $z$ and its properties, the disentangler $u$ and its properties, and the entanglement renormalization layer $Z$. Arrows denote edge connections.}
    \label{gates}
\end{figure}

\begin{figure}
    \centering
    \includegraphics[width=10cm]{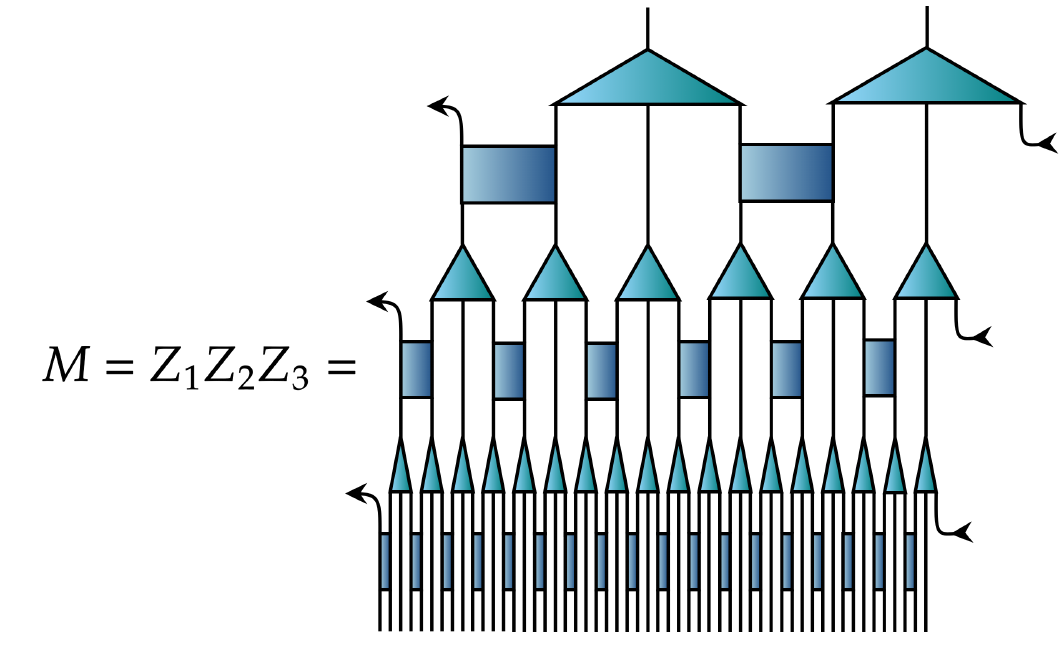}
    \caption{Diagram for MERA network $M$ consisting of three entanglement renormalization layers $Z_1$, $Z_2$, and $Z_3$.}
    \label{mera}
\end{figure}

Tensor networks have proven their efficiency in solving different challenging problems of quantum physics. They are useful not only for the analysis of many-body quantum systems \cite{orus2019tensor, orus2014practical, schollwock2011density} but also for the analysis of non-Markovian quantum dynamics \cite{luchnikov2019simulation, jorgensen2019exploiting} and benchmarking quantum circuits \cite{torlai2020quantum}. Some tensor networks impose specific constraints on subtensors. MERA construction requires the subtensors to be isometric~\cite{vidal2009entanglement, evenbly2009algorithms, evenbly2014algorithms, vidal2008class} and that enables it to describe critical many-body quantum systems. To introduce the notion of the entanglement renormalization let us consider a linear isometric mapping
\begin{eqnarray} \label{Z-operator}
Z:{\cal H}_1^{\otimes N}\rightarrow {\cal H}_2^{\otimes L},
\end{eqnarray}

\noindent where $N$ and $L$ are numbers of particles forming the input and output Hilbert spaces, respectively, and ${\cal H}_1$ and ${\cal H}_2$ are single particle Hilbert spaces with ${\rm dim}({\cal H}_i) = \chi_i$, $i=1,2$. Such a linear isometric mapping is used to effectively reduce the dimensionality of quantum systems if $\chi_1^N > \chi_2^L$. $Z$ provides a natural representation of the real-space renormalization group. For fixed orthonormal bases of input and output, the operator~\eqref{Z-operator} is given by a matrix $Z$ such that
$ZZ^\dagger$ is the identity matrix and $Z^\dagger Z$ is a projector. Since we consider an isometric mapping~\eqref{Z-operator} with $N$ input particles and $L$ output particles, we endow the matrix $Z$ by composed indices  \begin{eqnarray} \label{Z-tensor}
 Z_{\underbrace{j_1,\dots,j_L}_{\rm output} \, , \, \underbrace{i_1,\dots,i_N}_{\rm input}},
\end{eqnarray}

\noindent where each index $i_k$ takes $\chi_1$ different values and each index $j_k$ takes $\chi_2$ different values. The total number of elements in tensor~\eqref{Z-tensor} is $\chi_1^N\chi_2^L$. To be applicable to the cases when $N$ and $L$ scale proportionally as $N = 3^k L$, $k \in \mathbb{N}$, the tensor~\eqref{Z-tensor} is composed of two types of simpler submatrices (building blocks) from the Stiefel manifolds, namely, $u \in V_{\chi_1^2,\chi_1^2}$ and $z \in V_{\chi_1^3,\chi_2}$ called disentangler and elementary isometry, respectively. The disentangler satisfies
\begin{eqnarray}
u:{\cal H}_1^{\otimes 2}\rightarrow{\cal H}_1^{\otimes 2}, \quad u^\dagger u = u u^\dagger = I
\end{eqnarray}

\noindent and the elementary isometry satisfies
\begin{eqnarray}
z:{\cal H}_1^{\otimes 3}\rightarrow{\cal H}_2, \quad zz^\dagger = I,  \quad z^\dagger z = P,
\end{eqnarray}

\noindent where $P$ is a projector. The tensor diagram notation~\cite{orus2014practical} for $u$, $z$, and one entanglement renormalization layer with $L=N/3$ is given in Fig.~\ref{gates}. Merging entanglement renormalization layers into a single tensor network, we get the ternary MERA network~\cite{evenbly2009algorithms}, see Fig.~\ref{mera}. The role of base 3 in our description is not crucial; there exist, e.g., the binary MERA and modified binary MERA networks \cite{evenbly2013quantum}.

\begin{figure}
    \centering
    \includegraphics[width=14cm]{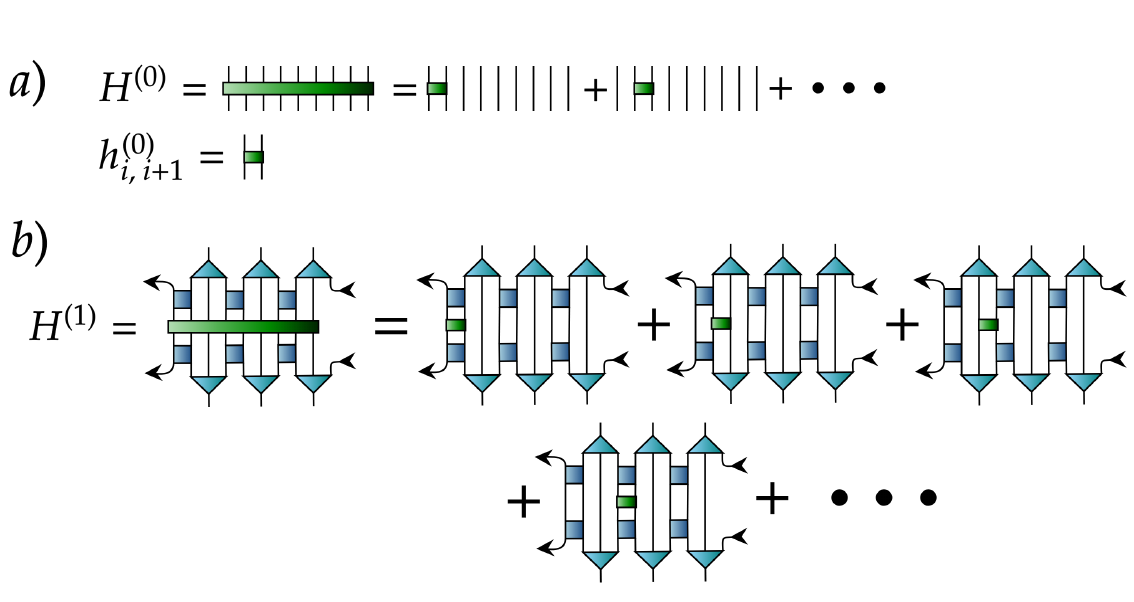}
    \caption{Tensor diagram for a local Hamiltonian of a spin chain (a). Tensor diagram for a renormalized local Hamiltonian of a spin chain (b).}
    \label{renorm_local_ham}
\end{figure}

\begin{figure}
    \centering
    \includegraphics[width=12cm]{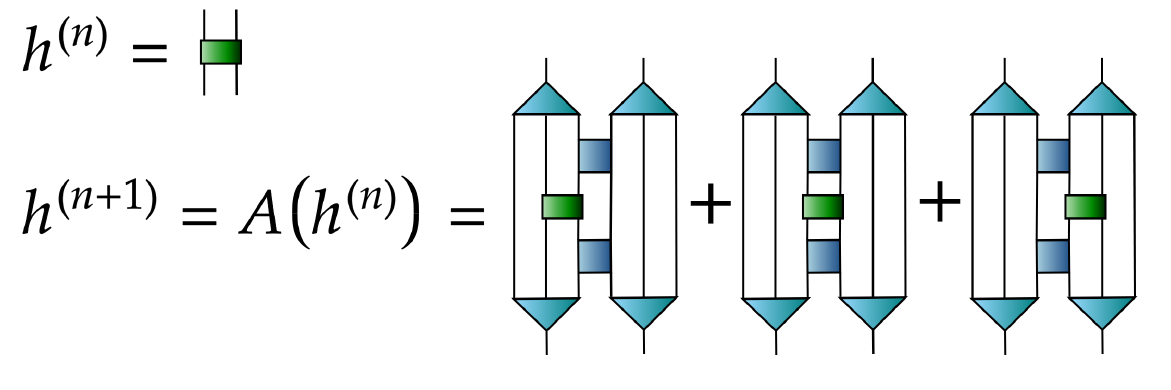}
    \caption{Renormalization of a local term in the system Hamiltonian by the ascending superoperator.}
    \label{renorm_local_term}
\end{figure}

\subsection{Renormalization theory for a local Hamiltonian} \label{local_ham_renorm}

The ideas of the Hamiltonian renormalization in Section~\ref{section-Hamiltonian-renorm} are directly applicable to MERA. Here we consider a class of local many-body Hamiltonians and its particular form for a spin chain with periodic boundary conditions, $H^{(0)} = h_{N,1}^{(0)} + \sum_{i=1}^{N-1} h_{i,i+1}^{(0)}$, where $N$ is a number of spins and $h_{i, i+1}^{(0)}$ is a local term acting on two neighboring sites with numbers $i$ and $i+1$. The superscript $(n)$ marks the order of renormalization (the number of entanglement renormalization layers used), so the superscript $(0)$ corresponds to the initial Hamiltonian and
\begin{eqnarray}
H^{(m+1)}=Z H^{(m)} Z^\dagger,
\end{eqnarray}

\noindent where the numbers of input and output particles are appropriate. The diagram for a local Hamiltonian $H^{(0)}$ and its first renormalization $H^{(1)}$ are depicted in Fig.~\ref{renorm_local_ham}. The remarkable property of isometry $Z$ is that the renormalized Hamiltonian $H^{(1)}$ is local too, i.e., $H^{(1)} = h_{N/3,1}^{(1)} + \sum_{i=1}^{N/3-1} h_{i,i+1}^{(1)}$, see Fig.~\ref{renorm_local_term}. In general we have
\begin{equation}
H^{(n)}=h^{(n)}_{N/3^n,1}+\sum_{i=1}^{N/3^n - 1 }h^{(n)}_{i,i+1},
\end{equation}

\noindent where the recurrence relation for $h_{i,i+1}^{(n)}$ is expressed through the ascending superoperator $A^{(n)}$ by
$h_{i,i+1}^{(n+1)} = A^{(n)} \left( h_{i,i+1}^{(n)} \right)$. The tensor diagram illustrating the ascending superoperator action is shown in Fig.~\ref{renorm_local_term}. For a chain of $N = 2\times 3^n$ spins, the $n$-th order renormalization results in the following Hamiltonian:
\begin{eqnarray} \label{H-n}
H^{(n)}=h^{(n)}_{1, 2} + h^{(n)}_{2, 1}.
\end{eqnarray}

The ground state energy of the spin chain per site equals
\begin{eqnarray}
E_g = \frac{1}{N} \min_{\ket{\phi},\{u, z\}} \bra{\phi} H^{(n)} \ket{\phi},
\label{ground_state_optimization}
\end{eqnarray}
where $\ket{\phi}$ is a trial state in the $n$-th order renormalized Hilbert space of small dimension, $\{u, z\}$ is the set of all disentanglers and elementary isometries. As all elements of the set $\{u, z\}$ and the trial state $\ket{\phi}$ are elements of the complex Stiefel manifold, we use the Riemannian optimization algorithms on these manifolds to solve the problem \eqref{ground_state_optimization}.

\begin{figure}
    \centering
    \includegraphics[width=17cm]{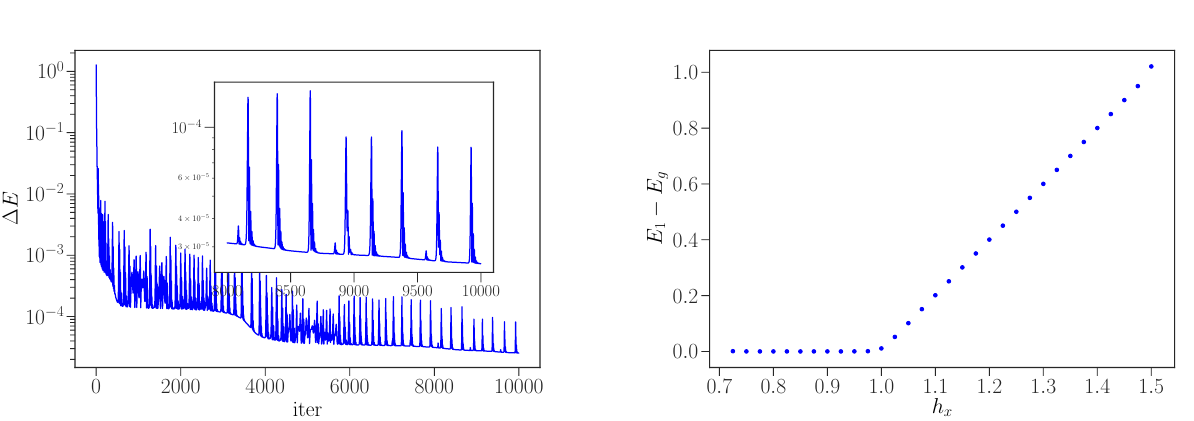}
    \caption{Left: Performance of the Riemannian {\scshape Adam} optimizer with the Euclidean metric on the complex Stiefel manifold and the SVD retraction in the evaluation of the ground state energy for the TFI model with $486$ spins. Right: Prediction for the energy gap per site in the TFI model vs the transverse field intensity provided by the same Riemannian {\scshape Adam} optimizer. Each dot corresponds to a result obtained after $10\,000$ iterations, with the step size exponentially decaying from $0.5$ to $0.03$. Other hyperparameters are standard ($\beta_1=0.9$, $\beta_2=0.999$, $\epsilon=10^{-8}$).}
    \label{energy_gap}
\end{figure}

\subsection{Results}
To benchmark the Riemannian optimization approach, we consider a one-dimensional  anti-ferromagnetic Ising model for spin-$\frac{1}{2}$ particles ($\chi_1 = 2$) in the transverse magnetic field (TFI model) with periodic boundary conditions. The system Hamiltonian reads
\begin{eqnarray} \label{TFI-Hamiltonian}
H = \sigma^z_N\sigma^z_1 + \sum_{i=1}^{N-1}\sigma^z_i\sigma^z_{i+1} + h_x\sum_{i=1}^N\sigma^x_i,
\label{ising_ham}
\end{eqnarray}
where $h_x$ is an external magnetic field directed along the $x$-axis, $\sigma^z_i$ and $\sigma^x_i$ are Pauli operators acting on a particle at site $i$. The model exhibits a quantum phase transition at $h_x=1$ in the thermodynamic limit ($N\rightarrow \infty$). Search for the ground state in the vicinity of the quantum phase transition is the most challenging for a numerical simulation, however, it is known that the conventional MERA optimization overcomes this difficulty~\cite{vidal2007entanglement}. Here we demonstrate that the Riemannian optimization approach successfully overcomes this difficulty too.

We consider the TFI model with $N = 2 \times 3^5 = 486$ spins and $h_x=1$. We apply $n=5$ entanglement renormalization layers to the Hamiltonian \eqref{ising_ham}. We bound the dimension $\chi_2$ of the local output Hilbert space ${\cal H}_2$ from above by $\chi_{\rm max} = 4$. In general, the greater $\chi_{\rm max}$ the higher accuracy is achieved. We use the TensorNetwork library \cite{roberts2019tensornetwork, ganahl2019tensornetwork} to design the ascending superoperators and perform the automatic differentiation described in Section~\ref{section-AD-and-RO}. Then we optimize the variational energy $E = \frac{1}{N} \bra{\phi} H^{(n)} \ket{\phi}$ by using the Riemannian {\scshape Adam} optimizer on the Stiefel manifolds. We perform $10\,000$ optimization steps with the step size exponentially decaying from $0.5$ to $0.03$. The code is available at \cite{QGOpt_repo}.

Since the exact value of the ground state energy per site is known for the TFI model with the external field $h_x=1$~\cite{He_2017}, $E_g = - 2 / N \sin(\pi/2N)$, we compare it with the output of the Riemannian {\scshape Adam} optimizer. The difference $\Delta E = E - E_g$ is shown in Fig.~\ref{energy_gap}. The achieved relative error for the ground state energy per site is of the order of $10^{-5}$.

Following the lines of Section~\ref{section-Hamiltonian-renorm}, to find the low-energy spectrum of the original Hamiltonian~\eqref{TFI-Hamiltonian} we minimize the cost function ${\rm tr}\left(H^{(n)}\right)$ with respect to disentanglers and elementary isometries by exploiting the Riemannian {\scshape Adam} optimizer. Diagonalization of the obtained renormalized Hamiltonian $H^{(n)}$ yields the low-energy spectrum. As an illustrative example we calculate the energy gap $E_1 - E_g$ between the first excited state and the ground state of the TFI Hamiltonian~\eqref{TFI-Hamiltonian} for different values of the transverse magnetic field $h_x$, see Fig.~\ref{energy_gap}. Thus, our optimizer predicts that the energy gap vanishes if $h_x<1$ and linearly grows with $h_x$ if $h_x\geq1$, which entirely agrees with the exact solution. The transition from a zero energy gap to a non-zero energy gap at $h_x = 1$ is a clear evidence of the quantum phase transition.

We emphasize two main advantages of the Riemannian optimization approach to the entanglement renormalization. First, this approach is easy in implementation in comparison with the conventional MERA optimization (based on iterative singular value decompositions) and sometimes achieves better accuracy~\cite{hauru-2020}. Second, the Riemannian optimization has a wide area of applications not limited to the low-energy physics of local Hamiltonians. Generality of the Riemannian optimization allows one to use it in different settings, e.g., in neural networks comprising MERA as a part. The library~\cite{QGOpt_repo} provides all necessary tools for accomplishing the Riemannian optimization on complex quantum architectures.

 \begin{figure}
     \centering
     \includegraphics[width=12cm]{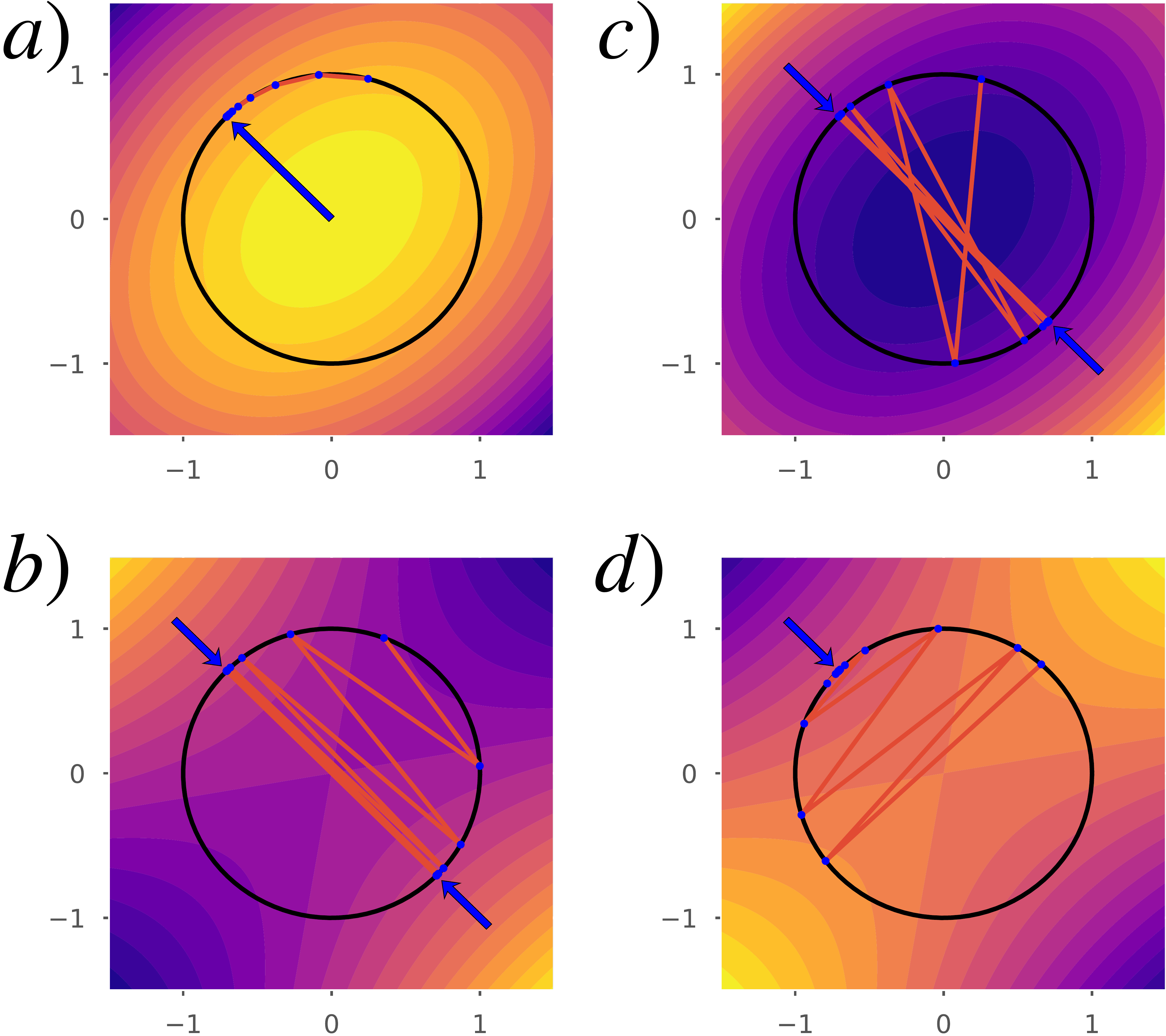}
     \caption{Convergence of the conventional MERA optimization algorithm while searching the minimum of $F(\omega) = \omega^{\dag} H \omega$ over orthogonal $2 \times 1$ matrices $\omega$. Horizontal and vertical exes correspond to the first and the second elements of $\omega$, respectively. Real symmetric $2\times 2$ matrix $H$ has eigenvalues $\lambda_1=-1$ and $\lambda_2=-2$ (a), $\lambda_1=-1$ and $\lambda_2=2$ (b), $\lambda_1=1$ and $\lambda_2=2$ (c), $\lambda_1=1$ and $\lambda_2=-2$ (d). Blue arrows mark limit points of the algorithm output.}
     \label{mera_quadratic_opt}
 \end{figure}

\subsection{Remarks on the conventional MERA optimization algorithm}\label{MERA_optimizer}

In Fig.~\ref{methods_comparison_stiefel}, we have already compared performances of different Riemannian optimization algorithms on the Stiefel manifold with the conventional MERA optimization algorithm. Here we discuss in more detail the latter optimization algorithm, which has been the main approach to the entanglement renormalization since 2009~\cite{evenbly2009algorithms, evenbly2014algorithms}. The main procedure within this algorithm solves the problem
\begin{equation} \label{linear_energy}
\begin{aligned}
   &\Min \quad f(\omega) := 2 {\rm Re \, tr}\left(\Gamma \omega\right) \\
   &\text{subject to}\quad \omega\in V_{n, m}, \\
\end{aligned}
\end{equation}

\noindent where $\omega$ is an isometric $n\times m$ matrix from the Stiefel manifold $V_{n, m}$, $m\leq n$, and $\Gamma$ is an arbitrary $m\times n$ matrix. The problem \eqref{linear_energy} has an exact solution
\begin{equation}  \label{mera_opt_step}
 \omega_{\rm opt} = - WU^{\dag},
\end{equation}

\noindent where $U$ and $W$ are matrices in the singular value decomposition $\Gamma = U\Lambda W^{\dag}$. Note that $f(\omega_{\rm opt}) = - 2 \sum_{i=1}^m\lambda_i$, where $\{\lambda_i\}_{i=1}^{m}$ is a set of singular values for $\Gamma$. The generalized algorithm for optimization of an arbitrary real-valued function $F(\omega)$ is performed by iterating the following three steps until convergence:
 \begin{eqnarray}
 &&G_{t+1} = \nabla F(\omega_t) \ \text{(compute the gradient)}, \nonumber\\
 && W_{t+1}\Lambda_{t+1} U_{t+1}^{\dag} = G_{t+1} \ \text{(compute the singular value decomposition of the gradient)}, \nonumber\\
 &&\omega_{t+1} = -W_{t+1}U_{t+1}^{\dag} \ \text{(compute a new isometric matrix for next iteration).}
 \end{eqnarray}
\noindent This algorithm can be seen as a first-order optimization algorithm because it uses the gradient. However, this algorithm converges to the optimal point for a very restricted set of problems only. Namely, this algorithm converges to the minimum of a negative-definite quadratic form, which is artificially guaranteed in the entanglement renormalization procedure by adding a significantly negative part to the Hamiltonian.

To illustrate the limits of this algorithm, we explore its performance while optimizing a quadratic form $F(\omega) = \omega^{\dag} H \omega$ in the simple case of orthogonal matrices $\omega$ with $n=2$ and $m=1$ and real symmetric $2\times 2$ matrix $H$. As $\omega$ is merely a two-component unit vector, the Stiefel manifold is the unit circle in this case. Fig.~\ref{mera_quadratic_opt} illustrates that the algorithm outputs a point $\omega_{\ast}$, which is an eigenvector of $H$ corresponding to the maximum absolute eigenvalue. If the quadratic form $\omega^{\dag} H \omega$ is negative-definite, then the algorithm converges to an eigenvector corresponding to the minimal eigenvalue of $H$. However, the algorithm fails in finding minimum of a general quadratic form $\omega^{\dag} H \omega$ as Figs.~\ref{mera_quadratic_opt}(b) and \ref{mera_quadratic_opt}(c) suggest. In contrast, the Riemannian optimization for the entanglement renormalization ansatz is free of such drawbacks.

\section{Riemannian optimization on the cone of positive-definite matrices} \label{section-cone-Spp}

\begin{figure}
    \centering
    \includegraphics[width=12cm]{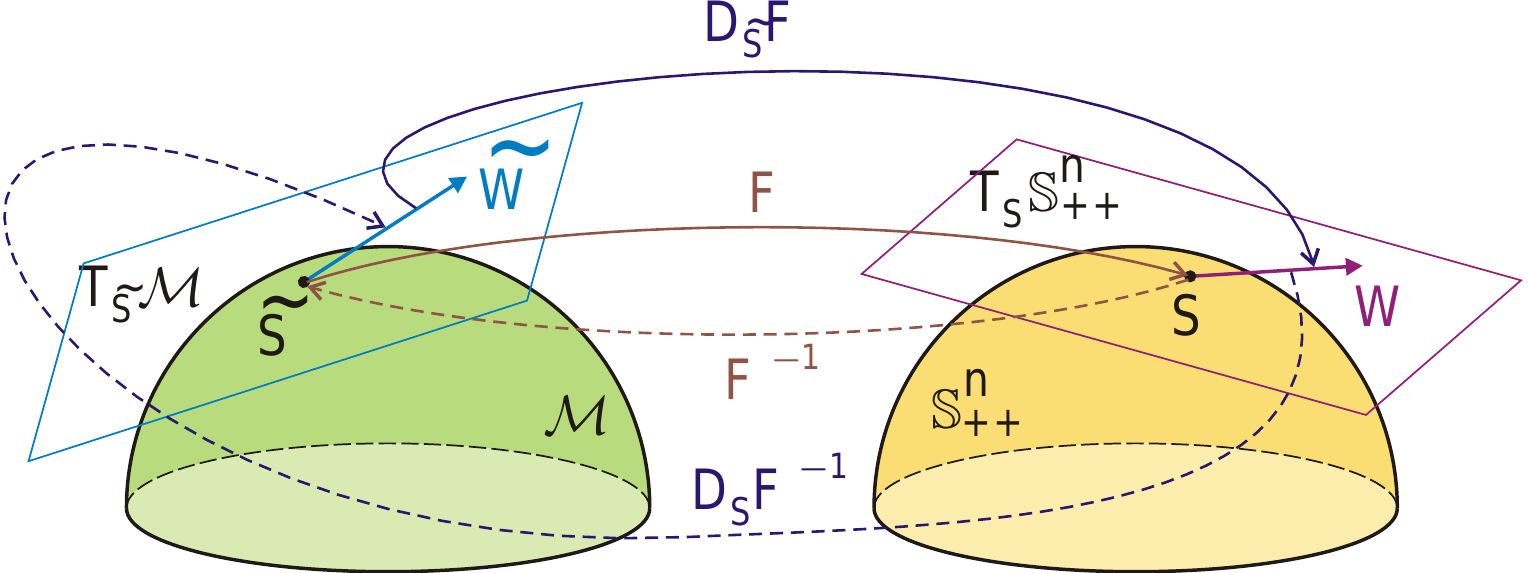}
    \caption{Relation between the Riemannian manifold ${\cal M}$ and the cone $\mathbb{S}_{++}^n$ of positive-definite matrices. For instance, ${\cal M}$ can be a manifold of Hermitian matrices $H$ (then $S = e^{H}$) or a manifold of lower triangular matrices $L$ with real and positive diagonal entries (then $S = LL^{\dag}$). In general, $F: {\cal M} \to \mathbb{S}_{++}^n$ is a smooth function and  $F^{-1}: \mathbb{S}_{++}^n \to {\cal M}$ is its smooth inverse. For a given positive-definite matrix $S\in \bS_{++}$ we denote $\w{S} = F^{-1}(S)$ its pre-image in $\bM$ (e.g., $\w{S}=H$ or $\w{S}=L$). Tangent vectors $W$ (in the tangent plane $T_S\bS_{++}^n$) and $\w{W}$ (in the tangent plane $T_{\w{S}}\bM$) are connected via the differentials $D_{\tilde{S}}F:T_{\tilde{S}}{\cal M} \rightarrow T_{S}\bS_{++}^n$ and $D_{\w{S}}F$ and $D_S F^{-1}: T_{S}\bS_{++}^n \rightarrow T_{\tilde{S}}{\cal M}$. }
    \label{fig:M-to-Spp}
\end{figure}

Any positive-definite matrix $S$ from the cone $\mathbb{S}_{++}^n$ defines the density operator $\rho = S / {\rm tr}[S]$. This fact enables optimization over the set of full-rank density matrices provided one can perform optimization on the cone $\mathbb{S}_{++}^n$ of positive-definite matrices. The cone $\mathbb{S}_{++}^n$ can be parameterized in many different ways (see, e.g.,~\cite{ilin-2018}). Here we consider two convenient ways to do that, namely, the exponential representation $S = e^{H}$, where $H$ is a Hermitian matrix, and the Cholesky decomposition $S = LL^{\dag}$, where $L$ is an $n \times n$ lower triangular matrix with real and positive diagonal entries. The both representations allow us to exactly derive formulas for geodesics and parallel transport within the corresponding metrics because the relation between $S$ on one side and $H$ or $L$ on the other side is a diffeomorphism. This means that there exist a smooth function $F: {\cal M} \to \mathbb{S}_{++}^n$ and its smooth inverse $F^{-1}: \mathbb{S}_{++}^n \to {\cal M}$ that maps $\mathbb{S}_{++}^n$ to a differentiable manifold ${\cal M}$, see Fig.~\ref{fig:M-to-Spp}. In our cases, ${\cal M}$ is composed of either Hermitian matrices $H$ or lower triangular matrices $L$ with real and positive diagonal entries. In the latter case, function $F$ is unique. Similarly, there exists an isomorphism between tangent bundles $T\mathbb{S}_{++}^n$ and $T{\cal M}$.

To unify and simplify the notation we will refer to all objects
associated with ${\cal M}$ and $T{\cal M}$ by adding symbol
$\widetilde{ }\ $ to the corresponding objects in
$\mathbb{S}_{++}^n$ and $T\mathbb{S}_{++}^n$, respectively, see
Fig.~\ref{fig:M-to-Spp}. For instance, for a given
positive-definite matrix $S\in \bS_{++}$ we denote $\w{S} =
F^{-1}(S)$ its pre-image in $\bM$. Analogously, for a tangent
vector $W\in T_S\bS_{++}^n$ we denote $\w{W}$ the corresponding
tangent vector in $T_{\w{S}}\bM$, which is connected with $W$ via
the differential $D_{\tilde{S}}F:T_{\tilde{S}}{\cal M}\rightarrow
T_{S}\bS_{++}^n$. We use the introduced tilda-notation for other
objects too, e.g., for exponential maps.

A Riemannian metric $\widetilde{g}$ on ${\cal M}$ induces some Riemannian metric $g$ on $\mathbb{S}_{++}^n$ and corresponding definitions of the exponential map and the parallel transport. The general scheme is as follows:
\begin{itemize}

\item For a given manifold $\bM$ and a diffeomorphism $F: {\cal M} \rightarrow \bS_{++}^n$ calculate the differentials $D_{\w{S}} F:T_{\w{S}}{\cal M}\rightarrow T_{S}\bS_{++}^n$ and $D_{S} F^{-1}:T_{S}\bS_{++}^n \rightarrow T_{\w{S}}{\cal M}$.

\item Define a standard Riemannian metric $\w{g}_{\w{S}}(\w{W}, \w{V})$ on $\bM$ and establish the exponential map $\w{\rm Exp}_{\w{S}}(\w{W})$ and the parallel transport $\w{\tau}_{\w{S},\w{W}} (\w{Q})$ of a vector $\w{Q}$ along a geodesic starting at point $\w{S}$ and aligned with the direction vector ${\w{W}}$ at point $\w{S}$.

\item Push forward all the introduced maps to $\bS_{++}^n$ by using $F$. This means that the induced Riemannian metric $g_S$ on $\bS_{++}^n$ is
\begin{equation} \label{g-M-S}
g_S (W, V) = \w{g}_{F^{-1}(S)} \left( D_S F^{-1}(W), D_S F^{-1} (V) \right) = \w{g}_{\w{S}} (\w{W}, \w{V}),
\end{equation}

\noindent the exponential map is
\begin{equation} \label{Exp-push-forward}
{\rm Exp}_{S} (W) = F \left( \w{\rm Exp}_{F^{-1}(S)} \big(D_S F^{-1}(W) \big) \right),
\end{equation}

\noindent and the parallel transport is
\begin{equation} \label{tau-push-forward}
\tau_{S,W}(Q) = D_{\w{\rm Exp}_{F^{-1}(S)} \big( D_S F^{-1}(W) \big)} F \left (\w{\tau}_{F^{-1}(S),D_S F^{-1}(W)} \big(D_S F^{-1}(Q) \big) \right).
\end{equation}

\end{itemize}

In Sections \ref{section-log-Euclidean} and \ref{section-log-Cholesky} we specify the above formulae in two cases: (i) ${\cal M}$ is a manifold of Hermitian matrices with the standard Euclidean metric $(V,W) = 2{\rm Re \, tr}(V^{\dag}W)$ and $F(H) = e^{H}$, which induces a so-called log-Euclidean Riemannian geometry on $\bS_{++}^n$, Refs.~\cite{Li2008LogEuclidean,huang2015log}; (ii) ${\cal M}$ is a manifold of lower triangular matrices with strictly positive diagonal elements that has a particular non-Euclidean metric and $F(L) = LL^{\dag}$, which induces a so-called log-Cholesky Riemannian geometry on $\bS_{++}^n$, Ref.~\cite{lin2019riemannian}.

\subsection{Log-Euclidean Riemannian geometry} \label{section-log-Euclidean}

Physical meaning of the exponential representation is that a density matrix $\rho = S / {\rm tr}[S]$, where $S = e^H$, can be thought of as an equilibrium thermal state with the effective Hermitian Hamiltonian $-kTH$, where $k$ is the Boltzmann constant and $T$ is a non-zero temperature. A one-parameter curve $S(t)$ in the cone $\bS_{++}^n$ of positive-definite matrices has a corresponding one-parameter preimage $H(t)$ in the space of Hermitian matrices ${\bM} = \{ M \in \mathbb{C}^{n \times n} | M = M^{\dag}\}$, i.e., in the space of Hamiltonians (up to an irrelevant factor $-kT$). Our goal is to find the exponential map ${\rm Exp}_S(W)$, the vector transport $\tau_{S,W}$, and the Riemannian gradient $\nabla_R f(S)$ for a function $f: \bS_{++}^n \to \mathbb{R}$ by using some simpler expressions for ${\bM}$ and $T_H{\bM}$.

To begin with, we follow the lines of Refs.~\cite{Li2008LogEuclidean,huang2015log} and derive differentials $D_{\w{S}}F$ and $D_S F^{-1}$, where $\w{S} \equiv H$. As
\begin{equation}
S(t) = e^{H(t)} = \lim\limits_{\epsilon \to 0} \big( I + \epsilon H(t) \big)^{1/\epsilon} = \lim\limits_{\epsilon \to 0} \prod_{k=1}^{1/\epsilon} \big( I + \epsilon H(t) \big),
\end{equation}

\noindent the tangent vector reads

\begin{eqnarray}
\frac{d S(t)}{d t} &=& \lim\limits_{\epsilon \to 0} \sum\limits_{l=1}^{1/\epsilon} \left[ \prod_{k=1}^{l-1} \big( I + \epsilon H(t) \big) \right] \epsilon \frac{dH(t)}{dt} \left[ \prod_{k=l+1}^{1/\epsilon} \big( I + \epsilon H(t) \big) \right] \nonumber\\
&=&  \lim\limits_{\epsilon \to 0} \sum\limits_{l=1}^{1/\epsilon} \big( I + \epsilon H(t) \big)^{l-1} \epsilon \frac{dH(t)}{dt} \big( I + \epsilon H(t) \big)^{1/\epsilon - l}.
\end{eqnarray}

\noindent Denoting $l\epsilon$ by $\tau$ and using the relation $\lim\limits_{\epsilon \to 0} \sum\limits_{l=1}^{1/\epsilon} \epsilon \, y(l\epsilon) = \int_0^1 y(\tau) d\tau$ for a function $y$ with domain $[0,1]$, we get
\begin{equation}
\frac{d S(t)}{d t} = \int\limits_{0}^1 d\tau e^{\tau H(t)} \frac{dH(t)}{dt} e^{(1-\tau) H(t)}.
\end{equation}

\noindent The spectral decomposition $H(t) = \sum_{i=1}^n \lambda_i(t) \ket{\psi_i(t)} \bra{\psi_i(t)}$ enables us to calculate the obtained integral explicitly, namely,
\begin{eqnarray}
\frac{d S(t)}{d t} & = & \sum\limits_{i,j=1}^n \int\limits_{0}^1 d\tau e^{\tau \lambda_i(t)} \ket{\psi_i(t)} \bra{\psi_i(t)} \frac{dH(t)}{dt} \ket{\psi_j(t)} \bra{\psi_j(t)} e^{(1-\tau) \lambda_j(t)} \nonumber\\
& = & \sum\limits_{i,j=1}^n G_{ij}(t) \ket{\psi_i(t)} \bra{\psi_i(t)} \frac{dH(t)}{dt} \ket{\psi_j(t)} \bra{\psi_j(t)},
\end{eqnarray}

\noindent where

\begin{equation} \label{f-matrix}
G_{ij}(t) = \int\limits_{0}^1 e^{\tau [\lambda_i(t) - \lambda_j(t)] + \lambda_j(t)}  d\tau = \left\{ \begin{array}{ll}
   \frac{e^{\lambda_i(t)} - e^{\lambda_j(t)}}{\lambda_i(t) - \lambda_j(t)}  & \text{if~} \lambda_i(t) \neq \lambda_j(t),  \\
   e^{\lambda_i(t)}  & \text{if~} \lambda_i(t) = \lambda_j(t).
\end{array} \right.
\end{equation}

\noindent The derived relation between operators $\frac{dS(t)}{dt}$ and $\frac{dH(t)}{dt}$ takes a simpler form in the index-free matrix representation in some fixed orthonormal basis $\{\ket{i}\}_{i=1}^n$, namely,
\begin{equation} \label{derivatives-S++-through-Hermitian}
\frac{d S(t)}{d t} = U(t) \left( G(t) \circ \Big( U^\dagger(t) \frac{dH(t)}{dt} U(t) \Big) \right) U^\dagger(t),
\end{equation}

\noindent where $U(t) = \sum_{i=1}^n \ket{\psi_i(t)}\bra{i}$ is a unitary transition matrix diagonalizing $H(t)$, $G(t) = \sum_{i,j=1}^n f_{ij}(t) \ket{i}\bra{j}$ is a symmetric matrix with elements $G_{ij}(t)$, and $\circ$ denotes the Hadamard (element-wise) product. Recalling the notation $\w{S} \equiv H$, we observe that Eq.~\eqref{derivatives-S++-through-Hermitian} explicitly defines the sought differentials $D_{\w{S}} F:T_{\w{S}}{\cal M}\rightarrow T_{S}\bS_{++}^n$ and $D_{S} F^{-1}:T_{S}\bS_{++}^n \rightarrow T_{\w{S}}{\cal M}$ via general expressions
\begin{eqnarray}
&& W = D_{\w{S}} F (\w{W}) = U_{\w{S}} \left (G_{\w{S}} \circ (U_{\w{S}}^\dagger \w{W} U_{\w{S}}) \right ) U_{\w{S}}^\dagger, \\
&& \w{W} = D_{S} F^{-1} (W) = U_S \left( G_{\w{S}}^{\circ -1} \circ (U_S^\dagger W U_S) \right ) U_S^\dagger,
\end{eqnarray}
where $U_S$ and $U_{\w{S}}$ are coincident unitary matrices diagonalizing both $S$ and $\w{S} \equiv H$, $G_{\w{S}}$ is a matrix of elements~\eqref{f-matrix} with $\{\lambda_i\}_{i=1}^n$ being the spectrum of $\w{S}$, and $G_{\w{S}}^{\circ -1} \equiv G_{F^{-1}(S)}^{\circ -1}$ is the Hadamard (element-wise) inverse of $G_{\w{S}}$, which is well defined because all the elements \eqref{f-matrix} are non-zero.

Then we equip the manifold ${\cal M}$ with the standard Euclidean metric $\w{g}_{\w{S}}(\w{W}, \w{V}) = 2 {\rm Re} \, {\rm tr}(\w{W}^\dagger \w{V})$, which makes the exponential map $\widetilde{\rm Exp}$ and the parallel transport $\widetilde{\tau}$ trivial, namely,
\begin{equation}
\widetilde{\rm Exp}_{\w{S}}(\widetilde{W}) = \w{S} +
\widetilde{W}, \qquad \widetilde{\tau}_{\w{S},\w{V}}(\w{W}) =
\w{W}.
\end{equation}

\noindent We pushforward the obtained formulae for the exponential map and the parallel transport on ${\cal M}$ to those on $\bS_{++}^n$ by using Eqs.~\eqref{Exp-push-forward} and~\eqref{tau-push-forward}. The induced Riemannian metric $g_S(W,V) = g_{\w{S}}(\w{W},\w{V})$ in $\bS_{++}^n$ is called log-Euclidean because $\w{S} = \log(S)$.

To get the Riemannian gradient $\nabla_R f(S)$ for a function $f: \bS_{++}^n \to \mathbb{R}$, we use Defenition~\ref{riemannian_grad_def},  property ${\rm tr}[A(G \circ B)] = {\rm tr}[(G^T \circ A)B]$, and symmetry $G_{\w{S}}=G_{\w{S}}^T$. This yields
\begin{equation}
\nabla_R f(S) = U_S \left[ G_{\w{S}} \circ \left( U_S^\dagger \frac{\nabla f(S) + \big(\nabla f(S)\big)^\dag}{2} U_S \right) \circ G_{\w{S}} \right] U_S^\dagger,
\end{equation}

\noindent where $H \equiv \w{S} = F^{-1}(S) = \log(S)$.

\subsection{Log-Cholesky Riemannian geometry} \label{section-log-Cholesky}

In this section, we use results of the recent paper~\cite{lin2019riemannian} and derive on their basis the Riemannian gradient in another metric. We consider a manifold $\bM$ of lower triangular matrices with real and positive diagonal elements, so
$F(\w{S}) = \w{S} \w{S}^\dagger\in \bS_{++}^n$ and $\w{S} = F^{-1}(S)$ is the unique lower Cholesky factor of a matrix $S\in \bS_{++}^n$.

For the convenience of the reader, in what follows we use the following notation. By $\half{X}$ we denote the lower-triangular part of a matrix $X$ with diagonal elements halved. So $X = \half{X} + \half{X^T}^{T}$ for an arbitrary matrix $X$. We put $\low{X}$ for the strictly lower-triangular part of a matrix $X$ and $\diag{X}$ for its diagonal part, so that we have $X = \low{X} + \diag{X} + {\low{X^T}}^T$ and $\half{X} = \low{X} + \frac{1}{2}\diag{X}$ for any matrix $X$.

To find the Riemannian gradient, we first calculate differentials
\begin{eqnarray*}
&& W = D_{\w{S}} F (\w{W}) = \w{S} \w{W}^\dagger + \w{W} \w{S}^\dagger, \\
&& \w{W} = D_{S} F^{-1} (W) = \w{S} \left ( \w{S}^{-1} W (\w{S}^{-1})^{\dagger} \right)_{\frac{1}{2}}.
\end{eqnarray*}

\noindent Then we follow Ref.~\cite{lin2019riemannian} and equip $\bM$ with the following Riemannian metric:
\begin{equation} \label{R-metric-M}
\w{g}_{\w{S}}(\w{W}, \w{V}) = {\rm tr} \left [ \low{\w{W}}^{\dag}\low{\w{V}} + \diag{\w{W}}^{\dag} \diag{\w{S}}^{-2} \diag{\w{V}} \right],
\end{equation}

\noindent which induces the Riemannian metric $g_S(W,V) = g_{\w{S}}(\w{W},\w{V})$ in $\bS_{++}^n$ and is called log-Cholesky metric~\cite{lin2019riemannian} rather in analogy with log-Euclidean metric though no logarithm is used in the expression for the Cholesky lower factor $\w{S} = F^{-1}(S)$.

The exponential map and the parallel transport have the following form~\cite{lin2019riemannian}:
\begin{eqnarray}
&& \label{exp-map-M-cholesky} \w{\rm Exp}_{\w{S}}(\w{W}) = \low{\w{S}} + \low{\w{W}} + \diag{\w{S}} \exp{\left (\diag{\w{W}} \diag{\w{S}}^{-1}\right )},\\
&& \w{\tau}_{\w{S}, \w{W}} (\w{Q}) = \low{\w{Q}} + \diag{\w{\rm Exp}_{\w{S}}( \w{W})}\diag{\w{S}}^{-1}\diag{\w{Q}}. \label{vector-transport-M}
\end{eqnarray}

\noindent We finish by pushing the relations \eqref{exp-map-M-cholesky}--\eqref{vector-transport-M} forward along the mappings $F(\w{S}) = \w{S} \w{S}^\dagger$ and $D_{\w{S}} F (\w{W}) = \w{S} \w{W}^\dagger + \w{W} \w{S}^\dagger$, see Eqs.~\eqref{g-M-S}--\eqref{tau-push-forward}.

Using Definition~\ref{riemannian_grad_def}, we finally derive the Riemannian gradient
\begin{eqnarray}
&& \nabla_R f(S) = \w{\nabla_R f(S)}  \, \w{S}^{\dag} + \w{S} \, \w{\nabla_R f(S)}^\dag, \\
&& \w{\nabla_R f(S)} = \left\lfloor \left( \nabla f(S) + \big( \nabla f(S) \big)^{\dag} \right) \w{S} \right\rfloor + \frac{1}{2} \diag{\w{S}}^2 \mathbb{D} \left\{\left( \nabla f(S) + \big( \nabla f(S) \big)^{\dag} \right) \w{S} + \w{S}^{\dag} \left( \nabla f(S) + \big( \nabla f(S) \big)^{\dag} \right) \right\}, \label{preimage-gradient-Cholesky}
\end{eqnarray}

\noindent where we have used notation $\w{S} = F^{-1}(S)$ for the lower Cholesky factor of $S$ and taken into account the fact that the preimage $\w{\nabla_R f(S)}$ must be a lower triangular matrix with real diagonal elements.

\subsection{Ground state search and other problems} \label{section-Spp-ground}

\begin{figure}
    \centering
    \includegraphics[width=17cm]{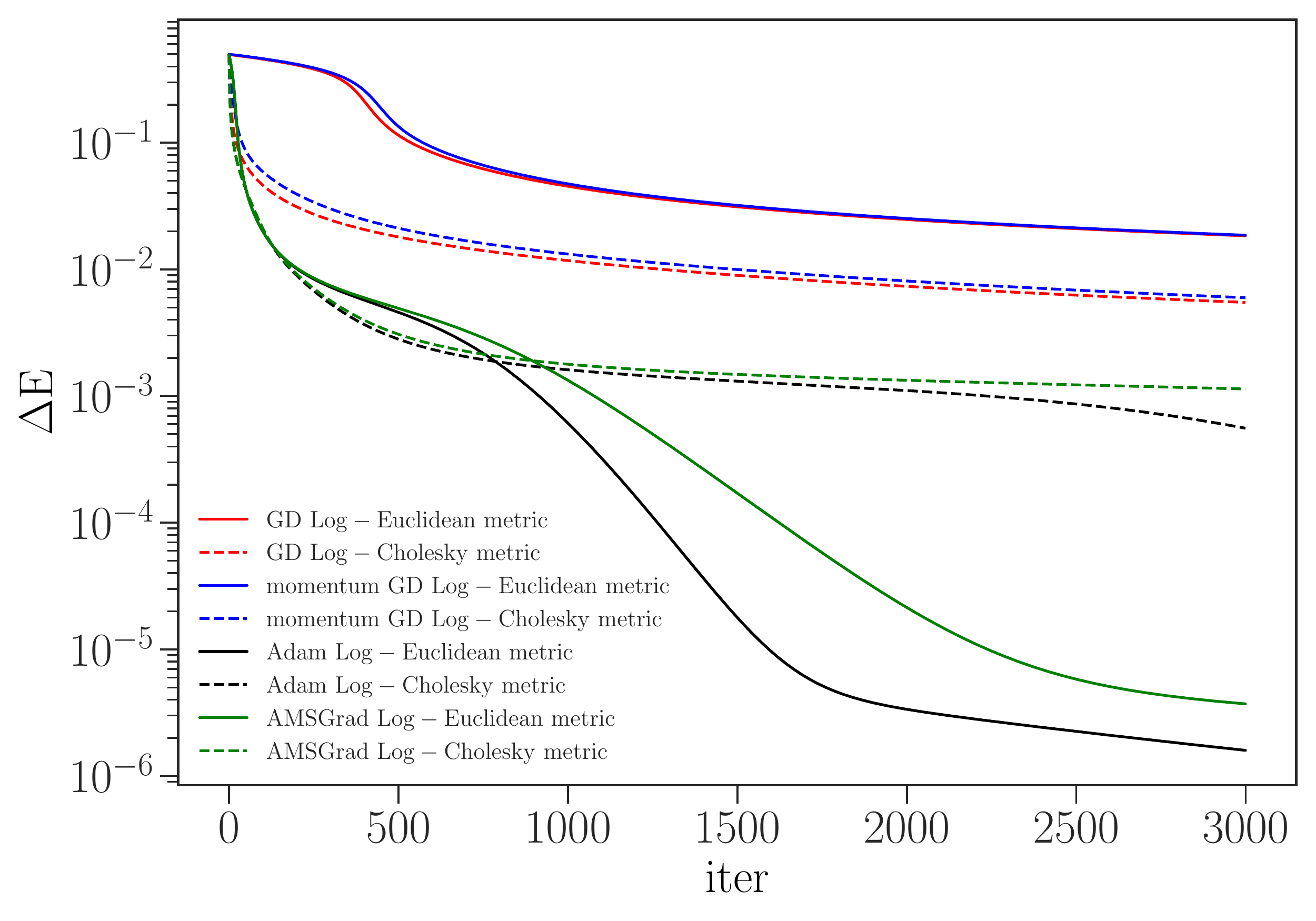}
    \caption{Difference $\Delta E = {\rm tr}\left(HS\right) / {\rm tr}\left(S\right) - E_{g}$ in solving of the problem \eqref{sdp_gs_search} vs iteration number for various Riemannian optimization algorithms. The optimization step size for all optimization methods is $\eta=0.5$. The dimension of the Hamiltonian matrix is $100 \times 100$. Other hyperparameters are standard for the algorithms ($\beta = 0.9$ for the GD with momentum; $\beta_1 = 0.9$, $\beta_2 = 0.999$, $\epsilon=10^{-8}$ for {\scshape Adam} and {\scshape AMSGradand}).}
    \label{methods_comparison_dens}
\end{figure}

To address performance of the Riemannian optimization on the cone of positive-definite matrices $\mathbb{S}_{++}^n$, we consider a quantum system in the Hilbert space of dimension $n=100$ and a random Hamiltonian $H$, whose ground state and ground energy are of interest. We benchmark the results of several Riemannian optimization algorithms with the actual ground energy for $H$, which is known at the stage of Hamiltonian generation but not accessible to optimization algorithms. Technically, the spectrum of $H$ is a sample from a uniform distribution of eigenvalues within the segment $[0, 1]$ and the eigenbasis for $H$ is obtained via the QR decomposition of a random matrix. The optimization problem is
\begin{equation} \label{sdp_gs_search}
\begin{aligned}
   &\Min \quad f(S) := \frac{{\rm tr}\left(HS\right)}{{\rm tr}\left(S\right)} \\
   &\text{subject to}\quad S\in \mathbb{S}^n_{++}, \\
\end{aligned}
\end{equation}

\noindent where $\rho = \frac{S}{{\rm tr}S}$ is a full-rank density matrix. The actual ground state $\rho_{g}$ is a rank-1 operator, which is a boundary point for the cone of positive-definite matrices. For this reason  $\inf_{S\in \mathbb{S}^n_{++}} \frac{{\rm tr}\left(HS\right)}{{\rm tr}\left(S\right)} = {\rm tr}[\rho_{g}H] = E_{g}$; however, one can only approach but not reach the infimum during the optimization process with a finite number of iterations. Thus, the optimization problem is challenging for Riemannian optimization algorithms on $\mathbb{S}_{++}^n$. We have tested several Riemannian optimizers including the Riemannian GD, the Riemannian GD with momentum, the Riemannian {\scshape Adam}, and the Riemannian  {\scshape AMSGrad} on the cone $\mathbb{S}_{++}^n$ with both the log-Euclidean metric and the log-Cholesky metric. Performance of a given algorithm is evaluated by a difference between the energy ${\rm tr}(HS) / {\rm tr}(S)$ at current iteration with the actual ground state energy $E_{g}$. Comparison of performances for various Riemannian optimization algorithms is presented in Fig.~\ref{methods_comparison_dens}. Efficacy of the Riemannian {\scshape Adam} and {\scshape AMSGrad} algorithms with the log-Euclidean metric allows one to find the ground state and its energy with precision $\sim 10^{-6}$ despite the fact that $\rho_{g}$ does not belong to the cone of positive-definite matrices. The code for implementation of algorithms is available at \cite{QGOpt_repo}.

Clearly, the considered problem is rather an illustration of the algorithm performance because it would be easier to find the ground state with the technique of Sec.~\ref{section-Hamiltonian-renorm}. However, there exist important problems of quantum information science, where the optimal density operator is mixed (not a pure state). For instance, the maximum of the coherent information $I_c(\Phi) = S(\Phi[\rho]) - S(\widetilde{\Phi}[\rho])$ gives the single-letter quantum capacity of the quantum channel $\Phi$, and this quantity is strictly greater than 0 only if $\rho$ is a mixed density operator ($\widetilde{\Phi}$ denotes the complementary channel), see Ref.~\cite{holevo2012quantum}. It was shown in some recent research papers~\cite{leditzky-2018,siddhu-2020,filippov-2021} that the two-letter quantum capacity can exceed the single-letter quantum capacity, which makes that optimization problem even more relevant. So we believe that the optimization on the cone of positive definite matrices has a number of possible applications.

\subsection{Tomography of quantum states}
\label{section-ST}

Quantum state tomography is an important statistical problem aimed at reconstructing an unknown density operator by processing results of a finite number $K$ of measurements performed on $K$ copies of the state \cite{d2003quantum,bogdanov2011statistical}. The higher the dimension of quantum system, the more challenging the problem. Linear reconstruction formulae suffer from possible non-positivity of the reconstructed operator, especially if the number of available copies $K$ is limited. In maximum likelihood reconstruction schemes, the optimization of the likelihood function is performed over a restricted set of positive-semidefinite matrices with unit trace. Here we show that the Riemannian optimization of the likelihood function over the cone of positive-definite operators successfully solves the tomography problem.

\begin{figure}
    \centering
    \includegraphics[width=14cm]{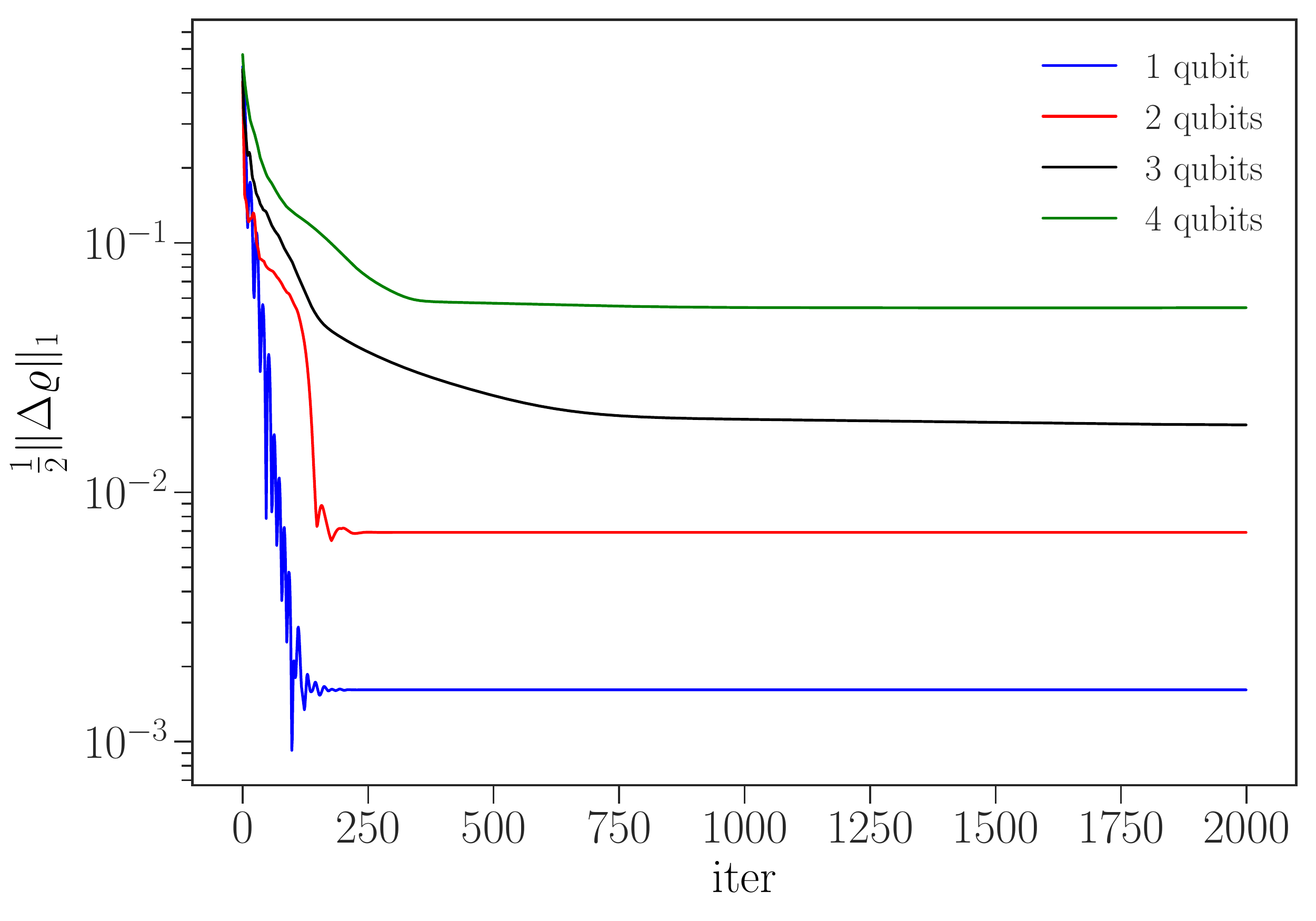}
    \caption{Performance of the Riemannian {\scshape Adam} optimizer based on the log-Cholesky metric in solving the tomographic reconstruction problem \eqref{tomography-optimization} for 1 to 4 qubits: Trace distance $\frac{1}{2}\| \Delta \rho \|_1 = \frac{1}{2}\| \rho_{\ast} - S/{\rm tr}(S)\|_1$ between the actual $N$-qubit state $\rho_{\ast}$ and the state $S/{\rm tr}(S)$ obtained after a fixed number of iterations in the optimization algorithm is implemented. Optimization step size $\eta=0.3$, the number of measurements $K = 5 \times 10^5$. Other hyperparameters are standard ($\beta_1 = 0.9$, $\beta_2 = 0.999$, $\epsilon=10^{-8}$).}
    \label{trace_dist}
\end{figure}

We consider general $N$-qubit states, where each qubit is measured individually in a symmetric informationally complete setup. This means that the measurement on a single qubit has 4 outcomes $\{\alpha\}_{\alpha=0}^3$ and is described by the tetrahedral positive operator valued measure (POVM) with effects~\cite{caves}
\begin{eqnarray}
&& M^{\alpha}_{\rm tetra} = \frac{1}{4}\left(I + \bm{s}^{\alpha}\bm{\sigma}\right), \quad \bm{s}^{\alpha} \in \mathbb{R}^3, \quad \bm{\sigma} = \left(\sigma_x, \sigma_y, \sigma_z\right), \quad \alpha \in \{0, 1, 2, 3\},\nonumber\\
&&\bm{s}^0 = (0, 0, 1), \ \bm{s}^1 = \left(\frac{2\sqrt{2}}{3}, 0, -\frac{1}{3}\right), \ \bm{s}^2 = \left(-\frac{\sqrt{2}}{3}, \sqrt{\frac{2}{3}}, -\frac{1}{3}\right), \ \bm{s}^3 = \left(-\frac{\sqrt{2}}{3}, -\sqrt{\frac{2}{3}}, -\frac{1}{3}\right),
\end{eqnarray}

\noindent where $(\sigma_x, \sigma_y, \sigma_z)$ is the set of Pauli matrices. Measurement on an $N$-qubit system is given by $4^N$ POVM elements of the form
\begin{equation}
M_{\rm tetra}^{\alpha_1 \ldots \alpha_N} = M_{\rm tetra}^{\alpha_1}\otimes\dots\otimes M_{\rm tetra}^{\alpha_N}, \quad \alpha_1, \ldots, \alpha_N \in \{0, 1, 2, 3\}.
\end{equation}

The probability to observe a particular collective outcome $\alpha_1 \ldots \alpha_N$ is given by the Born rule, namely,
\begin{eqnarray}
P(\alpha_1,\dots,\alpha_N|\rho_{\ast}) = {\rm tr}\left(\rho_{\ast} M_{\rm tetra}^{\alpha_1,\dots,\alpha_N}\right),
\end{eqnarray}

\noindent where $\varrho_{\ast}$ is the actual density operator of the unknown state.

Given $K$ measurement outcomes $\{\alpha_1^i,\dots\alpha_N^i\}_{i=1}^K$, one can calculate the likelihood of their registration, i.e., the probability to observe all the outcomes provided the system is in some density operator $\rho$,
\begin{equation} \label{likelihood}
\prod_{i=1}^K P(\alpha_1^i,\dots,\alpha_N^i|\rho).
\end{equation}

\noindent Maximum in \eqref{likelihood} is attained at some
density operator $\varrho_{\max}$ which is the best estimate for
$\rho_{\ast}$. Note that the logarithmic likelihood
$\log\prod_{i=1}^K P(\alpha_1^i,\dots,\alpha_N^i|\rho)$ attains
its maximum value at the same state $\rho_{\max}$. Using the
parameterization $\rho = S/{\rm tr}(S)$, we end up with the
following optimization problem on the cone of positive-definite
matrices:
\begin{equation} \label{tomography-optimization}
\begin{aligned}
   &\Max \quad f(S) := \sum_{i=1}^K\log{\rm tr}\left(M^{\alpha_1^i \ldots \alpha_N^i}_{\rm tetra}\frac{S}{{\rm tr}S}\right) \\
   &\text{subject to}\quad S\in \mathbb{S}_{++}^{2^N}. \\
\end{aligned}
\end{equation}

Clearly, $\sup_{S\in \mathbb{S}_{++}^{2^N}} f(S) = f(\rho_{\max})$. Here we test performance of the Riemannian {\scshape Adam} optimizer on the manifold $\mathbb{S}_{++}^{2^N}$ with the log-Cholesky metric. The figure of merit is the trace distance $\frac{1}{2}\| \Delta \rho \|_1 = \frac{1}{2}\| \rho_{\ast} - S/{\rm tr}(S)\|_1$ between the actual state $\rho_{\ast}$ and the state $S/{\rm tr}(S)$ updated after each iteration of the optimization algorithm. Fig.~\ref{trace_dist} depicts results for a typical numerical experiment on $N=1,2,3,4$ qubits in a state described by a randomly chosen $2^N \times 2^N$ density operator $\rho_{\ast}$. If total number of measurement outcomes $K=5 \times 10^5$, then the reconstruction error varies from approximately 0.2\% to 7\% depending on the number of qubits. The greater $K$, the better the state reconstruction quality.

\section{Conclusions}
\label{section-concl}

Optimization over a specific manifold is a typical problem in quantum information science, from fundamental questions like compatibility of observables~\cite{heinosaari}, operational restrictions in general probabilistic theories~\cite{filippov2020operational}, and entanglement robustness~\cite{vidal-1999,ffk} to practical questions like evaluation of quantum channel capacities~\cite{nagaoka,filippov-romp} and learning an unknown quantum noise~\cite{luchnikov2020machine,guo2020tensor,krastanov2020unboxing}. The problem becomes really involved in the high-dimensional cases and the complex tensor networks composed of many elements to be optimized simultaneously.

In this paper, we advocated the Riemannian optimization and the automatic differentiation as efficient methods to solve optimization problems on the manifolds of unitary and isometric matrices. One of the main advantages of our approach was its universality that enabled one to optimize any functional, for instance, any tensor network composed of unitary and isometric tensors without resorting to ad hoc methods valid for specific architectures only (like the conventional MERA optimization). We supported this claim by solving some problems of optimal control in circuit implementation of quantum computation aimed at preparing a desired (entangled) state or decomposing a given quantum gate into simpler gates. We also demonstrated that the Riemannian optimization over the manifold of isometric matrices enables one to effectively study the low-energy physics of high-dimensional (multipartite) systems with no regard to the Hamiltonian complexity (structure). By comparing our results with the previously known ones in Fig.~\ref{methods_comparison_stiefel}, we observed that the Riemannian optimization on the complex Stiefel manifold of isometric matrices outperformed the conventional MERA optimization algorithm in accuracy of reproducing the low-energy spectrum.

As quantum channels are naturally parameterized by isometric
matrices via the Stinespring
dilation~\cite{stinespring1955positive,holevo2012quantum}, we
believe that the presented analysis can find further applications
in optimization over the set of quantum channels. The need in such
an optimization was pointed out in
Refs.~\cite{luchnikov2020machine,krastanov2020unboxing,guo2020tensor},
where an unknown quantum noise is learned via the maximum
likelihood estimation for results of a sequence of projective
measurements. Another application of the Riemannian optimization
on the manifold of isometric matrices is a search for the optimal
code to transmit quantum information (quantum states) through
noisy communication lines, i.e., in evaluation of quantum capacity
of quantum channels. Last but not least is the problem of optimal
coherent control for a noisy quantum
system~\cite{dong2010quantum,morzhin2019minimal}, where the
methods of Riemannian optimization would be of great help.

We considered the cone of positive-definite matrices and used the exponential parameterization and the Cholesky decomposition to define the Riemannian metric on the cone and derive the corresponding Riemannian gradients. With the results obtained we demonstrated the efficacy of the Riemannian optimization on the cone to find the ground state of a high-dimensional Hamiltonian and to reconstruct an unknown multiqubit state via the maximum likelihood estimation. Possible applications of optimization over a cone of positive-definite matrices are numerous (study of coexistence of POVM effects and compatibility of POVMs~\cite{heinosaari}, quantification of quantum correlations~\cite{horodecki,filippov2019quantum}, falsifying the additivity hypothesis for the classical capacity of some quantum channels~\cite{hastings}, revealing the superadditivity of coherent information for quantum channels and estimating their quantum capacities~\cite{leditzky-2018,siddhu-2020,filippov-2021}, etc.).

To conclude, in this paper we showed a great potential of the Riemannian optimization and automatic differentiation to advance numerical analysis in problems of quantum physics and quantum information theory. In addition to the methodology we provided the computational package~\cite{QGOpt_repo} written on top of TensorFlow that serves as a toolbox for optimization of quantum architectures of arbitrary complexity.

\section{Acknowledgments}

I.A.L. and M.E.K. thank Henni Ouerdane, Jacob Biamonte, Stephen Vintskevich, and Emil Davletov for useful discussions. Results of Secs. VI B and VI C were obtained by I.A.L. with the support from the Russian Science Foundation Grant No. 19-71-10091. Results of Sec. VII were obtained by I.A.L. and S.N.F. with the support from the Foundation for the Advancement of Theoretical Physics and Mathematics ``BASIS'' under Project No. 19-1-2-66-1. Results of Sec. VIII B were obtained by S.N.F. in Valiev Institute of Physics and Technology of Russian Academy of Sciences, where S.N.F. was supported by Program No. 0066-2019-0005 of the Russian Ministry of Science and Higher Education.

\bibliographystyle{IEEEtran}
\bibliography{bibliography}

\end{document}